\documentclass[twocolumn]{aastex61}
\usepackage{hyperref}

\newcommand{\goes}{{\it GOES}}

\newcommand{\lasco}{{\it LASCO}}
\newcommand{\stereo}{{\it STEREO}}
\newcommand{\sdo}{{\it SDO}}

\accepted{for publication in ApJ on \today}
\shorttitle{On the factors determining solar flares}
\shortauthors{Baumgartner et al.}
\usepackage{threeparttablex, tablefootnote}

\begin{document}

\title{On the factors determining the eruptive character of solar flares}

\author{Christian Baumgartner}
\email{christian.baumgartner@edu.uni-graz.at}
\correspondingauthor{Julia K. Thalmann}
\author[0000-0001-8985-2549]{Julia K. Thalmann}
\email{julia.thalmann@uni-graz.at}
\author[0000-0003-2073-002X]{Astrid M. Veronig}
\email{astrid.veronig@uni-graz.at}
\affil{IGAM/Institute of Physics, University of Graz, A-8010 Graz, Austria}

\begin{abstract}
We investigated how the magnetic field in solar active regions (ARs) controls flare activity, i.e., whether a confined or eruptive flare occurs. We analyzed 44 flares of GOES class M5.0 and larger that occurred during 2011--2015. We used 3D potential magnetic field models to study their location (using the flare distance from the flux-weighted  AR center $d_{\mathrm{FC}}$) and the strength of the magnetic field in the corona above (via decay index $n$ and flux ratio). We also present a first systematic study of the orientation of the coronal magnetic field, using the orientation $\varphi$ of the flare-relevant polarity inversion line as a measure. We analyzed all quantities with respect to the size of the underlying dipole field, characterized by the distance between the opposite-polarity centers, $d_{\mathrm{PC}}$. Flares originating from underneath the AR dipole $(d_{\mathrm{FC}}/d_{\mathrm{PC}}<0.5$) tend to be eruptive if launched from compact ARs ($d_{\mathrm{PC}}\leq60$ Mm) and confined if launched from extended ARs. Flares ejected from the periphery of ARs ($d_{\mathrm{FC}}/d_{\mathrm{PC}}>0.5$) are predominantly eruptive. In confined events the flare-relevant field adjusts its orientation quickly to that of the underlying dipole with height ($\Delta\varphi\gtrsim40^\circ$ until the apex of the dipole field), in contrast to eruptive events where it changes more slowly with height. The critical height for torus instability, $h_{\mathrm{crit}}=h(n=1.5)$, discriminates best between confined ($h_{\mathrm{crit}}\gtrsim40$ Mm) and eruptive flares ($h_{\mathrm{crit}}\lesssim40$ Mm). It discriminates better than $\Delta\varphi$, implying that the decay of the confining field plays a stronger role than its orientation at different heights.
\end{abstract}

\keywords{Sun: activity -- Sun: magnetic fields -- Sun: flares -- Sun: coronal mass ejections (CMEs)}

\section{Introduction}
\label{S-Introduction} 

Solar flares often occur in combination with coronal mass ejections (CMEs). It is believed, that both phenomena are closely related and different manifestations of the same underlying physical processes \citep{Priest_Forbes_2002,2011LRSP....8....6S}. Flares associated with a CME are usually referred to as eruptive events, while flares that lack associated ejections of coronal plasma are called confined or "CME-less" events \citep{Svestka1986}. However, we lack a deeper understanding of the physical mechanisms behind flares and CMEs, which knowingly condition our Space Weather on Earth \citep[e.g.][]{Gosling1991}  To date, we suffer from moderate abilities to predict (i) when a solar flare will happen and (ii) whether or not it will develop an associated CME. 

\cite{2007ApJ...665.1428W} analyzed 8 large flares ($\ge$ \goes\ class X1.0) and found that confined/eruptive flares tend to occur closer to/farther from the flux-weighted center of their host AR. They also analyzed the horizontal magnetic flux that penetrates a vertical plane, aligned with the flare-relevant photospheric polarity inversion line (PIL). They defined a lower (1.0--1.1 $R_{\odot}$) and an upper (1.1--1.5 $R_{\odot}$) height regime within this plane, and the corresponding fluxes as $F_{\rm low}$ and $F_{\rm high}$. They evaluated the ``flux ratio'' as $F_{\rm low}/F_{\rm high}$, to quantify the relative strength of the horizontal magnetic field strength as a function of height in the solar corona. They found lower/higher flux ratios for confined/eruptive flares, indicating that a less strongly decaying magnetic field with height may have hindered the ejection of an associated CME.

Alternative measures to quantify the decay of the horizontal field as a function of coronal height have been used, e.g., in the form of the decay index, $n= -{\rm d}\,\ln\,B_{\rm hor}/{\rm d}\,\ln\,h$, with $B_{\rm hor}$ denoting the constraining (external) magnetic field and $h$ the height in the corona. It is based on the torus instability model by \cite{Kliem_Toeroek_2006} which predicts that a flux rope becomes unstable if the central axis reaches a height ($h_{\mathrm{crit}}$) at which $n$ reaches a critical value ($n_{\mathrm{crit}}$). In other words, a system becomes unstable when the overlying horizontal field strength decayes sufficiently fast. Based on magneto-hydrodynamic simulations involving different geometrical assumptions on the current channel mimicking the flux rope \citep[e.g.,][]{Toeroek_Kliem_2007,Fan_Gibson_2007,Demoulin2010,2015ApJ...814..126Z} and prominence observations \citep[e.g.,][]{Zuccarello2016}, $n_{\mathrm{crit}}$ is found in the range $\simeq$1.0--2.0. Despite this extended range of critical values $n_{\mathrm{crit}}=1.5$ is often used in observation-based studies that aim at assessing the likelihood of a CME to occur. As a common practice, the constraining magnetic field above the potentially unstable flux rope is approximated by the horizontal component of an associated potential (current-free) magnetic field model.

Recently, \cite{Wang2017} presented a statistical analysis of 60 flares. They found that in ARs that host eruptive flares, $h_{\mathrm{crit}}$ is reached at significantly lower coronal heights. They also found that $n$ increases monotonously with height in 86\% (84\%) of the eruptive/confined events studied, whereas the rest of the analyzed events exhibited a saddle-like profile \citep[see also, e.g.,][]{2011ApJ...732...87C}. They found the saddle-like profiles to be characteristic for an underlying ``multipolar'' photospheric magnetic flux concentration, i.e., when the line connecting the opposite-polarity centroids of the flaring AR did not pass through the flare-relevant PIL or was oriented almost parallel to it. Careful examination of the saddle-like $n(h)$ profiles also revealed significantly lower values of $n$ at the saddle bottoms in confined events.

In this paper we also investigate the large-scale magnetic field structure in flaring ARs, in context with the occurrence of eruptive or confined events. We study 44 large ($\ge$M5.0) flare events that occurred during the \sdo\ era, between January 2011 and December 2015. In contrast to earlier works, we seek to pin down the magnetic field related parameter that discriminates best between confined and eruptive events. Therefore, we assess the performance of different known measures that characterize the decay of the confining field above flaring ARs (decay index and flux ratio). 

Going beyond, we also analyze the change of orientation of the flare-related magnetic field as a function of height in the corona. Ideally, if the photospheric magnetic field of an AR is approximated by an East-West aligned symmetric potential (current-free) dipole field, the corresponding PIL would run in North-South direction, both at low as well as large heights. In reality, however, the orientation of PILs at low heights are often found to deviate strongly from such an idealized north-southward direction, but to transit towards the North-South direction with increasing coronal height \citep[see, e.g., Fig.\,2 of][]{2015RAA....15.1537J}. This motivates us to present here a first systematic study of the orientation of the flare-relevant PIL as a function of height in the corona, for a large number of flares.

In section \ref{S-Data} we describe our event sample as well as the data and methods used. The details of the analysis are described in \ref{S-Analysis}, the obtained results presented in \ref{S-Results} and discussed in \ref{S-Discussion}.

\section{Data and Modeling}
\label{S-Data}
    
\subsection{Event sample}\label{SS-EventSample}

We searched the \goes\ flare catalog\footnote{\url{www.ngdc.noaa.gov/stp/satellite/goes/index.html}} for flare events exceeding a peak soft X-ray (SXR) flux of $5\times10^{-4}~{\rm W\,m}^{-2}$ (i.e., flares of \goes\ class M5.0 and larger), that occurred between January 2011 and December 2015. From this set of events, we selected those that took place within 50 degrees from the solar disk center to avoid projection effects. 

To classify events as confined or eruptive, we consulted the \lasco\ CME catalog\footnote{\url{cdaw.gsfc.nasa.gov/CME_list/}} \citep{2009EM&P..104..295G} and additionally the CACTus\footnote{\url{sidc.oma.be/cactus/catalog.php}} \citep{2004A&A...425.1097R} and CORIMP CME lists\footnote{\url{alshamess.ifa.hawaii.edu/CORIMP/}}. We regarded a flare and a CME to be associated, if the back-extrapolated height-time profile of the CME (as deduced from \lasco-C2 images) 
agreed with the flare onset time of the \goes\ flare catalog.
In addition, the position angle of the CME had to agree with the quadrant on the Sun in which the flare occurred. Furthermore, a flare was classified as to be eruptive if a coronal EUV wave was observed\footnote{\url{http://aia.lmsal.com/AIA_Waves/index.html}}, since these large-scale disturbances are known to be generated by magnetic reconfiguration in the framework of an expanding CME \citep[for a review see, e.g.,][]{Warmuth2015}. Only one event (SOL20120510T04:18M5.7) occurred during a short period of non-coverage in the \lasco\ CME catalog. In order to be able to classify the flare as confined or eruptive, we visually inspected \stereo-B COR1 observations, in which the flare was observed above the north-west limb. The absence of a flare-associated CME, together with the missing of a flare-associated EUV wave\footnote{\url{http://solardemon.oma.be/old/}}, allowed us to classify this event as to be confined.
 
In total, we were able to unambiguously deduce the flare type for 44 events which fulfill the selection criteria above (see Table~\ref{tab:flare_list}). Out of these, 12 ($\sim$27\%) events were confined (7 M- and 5 X-flares) and 32 ($\sim$73\%) were eruptive (18 M- and 14 X-flares). 
In comparison to \cite{2017ApJ...834...56T}, we classify two events differently. Firstly, we classify event no.\ 20 (SOL20110213T17:38M6.6) as to be eruptive because we observe an associated CME in \lasco\ C2. In addition, a coronal EUV wave has been observed, as analyzed by \cite{2017IAUS..327..109L}. Secondly, we classify event no.\ 33 (SOL20131101T19:53M6.3) as to be confined, for the opposite reasons.

The 32 eruptive flares in our event sample originated from 18 different ARs, and the 12 confined events originated from 7 different ARs. That implies that some of the considered events originated from a single AR within several consecutive days. For instance, five large confined flares originated from NOAA~12192 in the course of four days. Similarly, e.g., NOAA~11283 (11429) produced four large eruptive flares during three (four) consecutive days. As a consequence, some of the ARs are over-represented (especially in our set of confined events), we believe that the continuous time evolution of the individual ARs over several days allows us to study the corona above as representative for the conditions yielding confined or eruptive flaring. We nevertheless discuss the possible implications of the over-representation of NOAA~12192 in the sample of confined events when presenting our results.

\begin{table*}[t]
\begin{ThreePartTable}
\begin{center}
\begin{TableNotes}
\item [a] Flare peak time. 
\item [b] NOAA AR number.
\item [c] Flare class.
\item [d] Flare type: (E)ruptive/(C)onfined.
\item [e] Heliographic flare position.
\item [f] Distance between flare site and flux-weighted AR center (see Sect.~\ref{SS-DistanceFromPolarityCenter} for details).
\item [g] Distance between opposite magnetic polarity centers of AR (see Sect.~\ref{SS-DistanceFromPolarityCenter} for details).
\item [h] Critical height for the onset of torus instability $h_{\mathrm{crit}}=h(n=1.5)$ (see Sect.~\ref{SS-FluxRatioDecayIndex} for details).
\item [i] Change of orientation of the flare-relevant PIL in the range $h/d_{\mathrm{PC}}=[0,0.5]$ (see Sect.~\ref{SS-PIL_InclinationTrend} for details).
\end{TableNotes}
\caption{Event list (Flares $\ge$M5.0 that occurred between January 2011 and December 2015.)}
\label{tab:flare_list}
\renewcommand\arraystretch{0.9}
\begin{longtable*}{|c|c|c|c|c|c|c|l|l|l|l|l|}
\hline
\multicolumn{1}{|c|}{\textbf{Event}} & %
\multicolumn{1}{c|}{\textbf{Date}} & %
\multicolumn{1}{c|}{\textbf{Time}\tnote{a}} & %
\multicolumn{1}{c|}{\textbf{NOAA}\tnote{b}} & %
\multicolumn{1}{c|}{\textbf{SXR}\tnote{c}} & %
\multicolumn{1}{c|}{\textbf{Type}\tnote{d}} & %
\multicolumn{1}{c|}{\textbf{Flare}\tnote{e}} & %
\multicolumn{1}{c|}{\textbf{Flux}} & %
\multicolumn{1}{c|}{\textbf{$d_{\mathrm{FC}}$}\tnote{f}} & %
\multicolumn{1}{c|}{\textbf{$d_{\mathrm{PC}}$}\tnote{g}} & %
\multicolumn{1}{c|}{\textbf{$h_{\mathrm{crit}}$}\tnote{h}} & %
\multicolumn{1}{c|}{\textbf{$\Delta\varphi$}\tnote{i}} \\
\multicolumn{1}{|c|}{no.} & %
\multicolumn{1}{c|}{} & %
\multicolumn{1}{c|}{} & %
\multicolumn{1}{c|}{} & %
\multicolumn{1}{c|}{\textbf{class}}  & %
\multicolumn{1}{c|}{}  & %
\multicolumn{1}{c|}{\textbf{position}}  & %
\multicolumn{1}{c|}{\textbf{ratio}}  & %
\multicolumn{1}{c|}{\textbf{(Mm)}} & %
\multicolumn{1}{c|}{\textbf{(Mm)}} & %
\multicolumn{1}{c|}{\textbf{(Mm)}} & %
\multicolumn{1}{c|}{\textbf{[deg]}} \\
\endfirsthead

\multicolumn{11}{r}%
{{\tablename\ \thetable{} -- continued from previous page}} \\
\hline
\multicolumn{1}{|c|}{{No.}} & %
\multicolumn{1}{c|}{{Date}} & %
\multicolumn{1}{c|}{{Time}\tnote{a}} & %
\multicolumn{1}{c|}{{NOAA}} & %
\multicolumn{1}{c|}{{SXR}} & %
\multicolumn{1}{c|}{{Type}} & %
\multicolumn{1}{c|}{{Flare}} & %
\multicolumn{1}{c|}{{Flux}} & %
\multicolumn{1}{c|}{{$d_{\mathrm{FC}}$}\tnote{b}} & %
\multicolumn{1}{c|}{{$d_{\mathrm{PC}}$}\tnote{b}} & %
\multicolumn{1}{c|}{{$h_{\mathrm{crit}}$}\tnote{c}} & %
\multicolumn{1}{c|}{{$\Delta\varphi$}\tnote{d}} \\

\multicolumn{1}{|c|}{} & %
\multicolumn{1}{c|}{} & %
\multicolumn{1}{c|}{} & %
\multicolumn{1}{c|}{} & %
\multicolumn{1}{c|}{\textbf{class}}  & %
\multicolumn{1}{c|}{\textbf{E/C}}  & %
\multicolumn{1}{c|}{\textbf{position}}  & %
\multicolumn{1}{c|}{\textbf{ratio}}  & %
\multicolumn{1}{c|}{\textbf{(Mm)}} & %
\multicolumn{1}{c|}{\textbf{(Mm)}} & %
\multicolumn{1}{c|}{\textbf{(Mm)}} & %
\multicolumn{1}{c|}{\textbf{[deg]}} \\
\hline
\endhead

\hline
\multicolumn{11}{r}{{Continued on next page}}\\
\endfoot

\hline \hline
\insertTableNotes
\endlastfoot

\hline
1 & 20110215 & 01:56 & 11158 & X2.2 & E & S20W10 & 41.5 & 6.0 & 56.8 & 33.0 & 37.1 \\
2 & 20110309 & 23:23 & 11166 & X1.5 & C & N08W09 & 19.6 & 24.1 & 68.8 & 49.1 & -- \\
3 & 20110906 & 22:20 & 11283 & X2.1 & E & N14W18 & 233.2 & 44.3 & 69.1 & 11.1 & -- \\
4 & 20110907 & 22:38 & 11283 & X1.8 & E & N14W28 & 72.2 & 31.3 & 61.0 & 11.2 & -- \\
5 & 20120307 & 00:24 & 11429 & X5.4 & E & N17E31 & 82.3 & 4.4 & 42.6 & 27.4 & 21.6 \\
6 & 20120307 & 01:14 & 11429 & X1.3 & E & N22E12 & 71.7 & 18.5 & 42.4 & 19.1 & 19.2 \\
7 & 20120712 & 16:49 & 11520 & X1.4 & E & S15W01 & 22.1 & 34.7 & 40.3 & 26.8 & 18.0 \\
8 & 20131105 & 22:12 & 11890 & X3.3 & E & S12E44 & 40.0 & 49.3 & 64.3 & 6.5 & -- \\
9 & 20131108 & 04:26 & 11890 & X1.1 & E & S12E13 & 54.1 & 63.8 & 91.1 & 16.1 & -- \\
10 & 20131110 & 05:14 & 11890 & X1.1 & E & S14W13 & 62.1 & 62.2 & 98.2 & 20.1 & -- \\
11 & 20140329 & 17:48 & 12017 & X1.0 & E & N10W32 & 41.7 & 46.5 & 47.4 & 11.0 & -- \\
12 & 20140910 & 17:45 & 12158 & X1.6 & E & N11E05 & 13.9 & 10.4 & 47.8 & 28.8 & 25.9 \\
13 & 20141022 & 14:28 & 12192 & X1.6 & C & S14E13 & 7.3 & 3.0 & 113.2 & 59.4 & 49.2 \\
14 & 20141024 & 21:41 & 12192 & X3.1 & C & S22W21 & 4.5 & 23.9 & 123.8 & 78.7 & 44.1 \\
15 & 20141025 & 17:08 & 12192 & X1.0 & C & S10W22 & 6.2 & 16.1 & 120.3 & 72.7 & 55.6 \\
16 & 20141026 & 10:56 & 12192 & X2.0 & C & S14W37 & 8.5 & 21.1 & 100.2 & 58.3 & 62.5 \\
17 & 20141107 & 17:26 & 12205 & X1.6 & E & N17E40 & 9.1 & 4.6 & 55.7 & 45.5 & 55.3 \\
18 & 20141220 & 00:28 & 12242 & X1.8 & E & S19W29 & 12.0 & 52.3 & 107.4 & 37.8 & 24.8 \\
19 & 20150311 & 16:22 & 12297 & X2.1 & E & S12E22 & 62.3 & 29.8 & 37.9 & 9.4 & 48.2 \\
20 & 20110213 & 17:38 & 11158 & M6.6 & E & S20E04 & 629.7 & 6.7 & 45.6 & 11.6 & 32.7 \\
21 & 20110730 & 02:09 & 11261 & M9.3 & C & N14E35 & 58.8 & 16.9 & 37.0 & 21.7 & 59.0 \\
22 & 20110803 & 23:38 & 11261 & M6.0 & E & N16W30 & 68.0 & 23.1 & 38.6 & 12.1 & 2.7 \\
23 & 20110804 & 03:57 & 11261 & M9.3 & E & N19W36 & 46.9 & 26.7 & 41.5 & 12.5 & 4.7 \\
24 & 20110906 & 01:50 & 11283 & M5.3 & E & N14W07 & 253.9 & 54.2 & 76.4 & 12.9 & -- \\
25 & 20110908 & 14:46 & 11283 & M6.7 & E & N14W40 & 56.7 & 38.3 & 72.8 & 11.1 & -- \\
26 & 20120123 & 03:59 & 11402 & M8.7 & E & N28W21 & 15.4 & 5.3 & 45.1 & 27.7 & 30.4 \\
27 & 20120309 & 03:53 & 11429 & M6.3 & E & N15W03 & 20.6 & 43.6 & 53.5 & 17.8 & 26.6 \\
28 & 20120310 & 17:44 & 11429 & M8.4 & E & N17W24 & 22.4 & 34.1 & 55.7 & 22.3 & 8.7 \\
29 & 20120510 & 04:18 & 11476 & M5.7 & C & N10E22 & 49.1 & 58.1 & 71.4& 68.1 & -- \\
30 & 20120702 & 10:52 & 11515 & M5.6 & E & S17E08 & 35.6 & 51.8 & 52.1 & 24.2 & -- \\
31 & 20130411 & 07:16 & 11719 & M6.5 & E & N09E12 & 16.8 & 46.2 & 43.6 & 29.1 & 5.5 \\
32 & 20131024 & 00:30 & 11877 & M9.3 & E & S09E10 & 63.6 & 52.7 & 81.1 & 17.7 & -- \\
33 & 20131101 & 19:53 & 11884 & M6.3 & C & S12E01 & 12.6 & 17.1 & 74.8 & 54.7 & 49.1 \\
34 & 20131103 & 05:22 & 11884 & M4.9 & C & S12W17 & 11.4 & 8.5 & 89.1 & 48.1 & 59.2\\
35 & 20131231 & 21:58 & 11936 & M6.4 & E & S15W36 & 19.4 & 30.8 & 61.8 & 32.9 & -- \\
36 & 20140418 & 13:03 & 12036 & M7.3 & E & S20W34 & 15.3 & 42.1 & 41.3 & 22.2 & 6.1 \\
37 & 20141022 & 01:59 & 12192 & M8.7 & C & S12E21 & 9.9 & 11.8 & 109.2 & 47.4 & 38.7 \\
38 & 20141204 & 18:25 & 12222 & M6.1 & C & S20W31 & 20.0 & 24.1 & 68.4 & 46.7 & 60.2 \\
39 & 20141217 & 04:51 & 12242 & M8.7 & E & S18E08 & 26.6 & 76.1 & 128.2 & 24.6 & -- \\
40 & 20150309 & 23:53 & 12297 & M5.8 & E & S19E46 & 39.4 & 31.3 & 44.2 & 20.2 & 54.4 \\
41 & 20150310 & 03:24 & 12297 & M5.1 & E & S15E39 & 38.4 & 33.9 & 42.8 & 11.6 &  --\\
42 & 20150622 & 18:23 & 12371 & M6.5 & E & N13W06 & 15.1 & 30.1 & 93.9 & 22.6 & 37.7 \\
43 & 20150625 & 08:16 & 12371 & M7.9 & E & N12W40 & 16.6 & 16.7 & 76.5 & 34.6 & 35.5 \\
44 & 20150928 & 14:58 & 12422 & M7.6 & C & S20W28 & 67.0 & 45.7 & 91.4 & 27.9 & -- \\
\end{longtable*}
\end{center}
\normalsize
\end{ThreePartTable}
\end{table*}

\subsection{Data and methods}
\label{SS-ObservationalData}

The magnetic field configuration in and above the flare ARs was investigated based on full-disk vector magnetic field observations from the {\it Solar Dynamics Observatory} \citep[\sdo;][]{2012SoPh..275....3P} Helioseismic and Magnetic Imager \citep[HMI;][]{2012SoPh..275..229S}. We used {\sf hmi.B\_720s} data \citep{2014SoPh..289.3483H} which provides the total field, inclination and azimuth on the entire solar disk. After disambiguation of the provided azimuth\footnote{\url{jsoc.stanford.edu/jsocwiki/FullDiskDisamb}}, we applied a de-projection \citep[following ][]{1990SoPh..126...21G} to the image-plane transverse and line-of-sight (LOS) magnetic field ($B^x_t$,$B^y_t$,$B_{\rm LOS}$) in order to obtain the local horizontal and vertical magnetic field components ($B^x_h$,$B^y_h$,$B_z$). The native (full-resolution) plate scale of the magnetic field data is $\simeq$\,0\farcs504 per pixel. For magnetic field modeling, we binned the data to a plate scale of 1\farcs01 per pixel, with the binning of the data being nearly magnetic flux preserving. 

For an extended area, covering the flare AR as well as its near quiet-Sun surrounding, we modeled the potential magnetic field in the corona above, based on the Fourier transformation method outlined in \cite{1981A&A...100..197A}. Here, we used $B_z$ of the last available magnetic field data prior to the flare onset as input. We modeled the magnetic field vertically with the same resolution as in horizontal direction (1\farcs008 per pixel), up to a height of $1.26~R_{\odot}$ above a photospheric level (using $n_z=256$ levels in our numerical setup).

We used high-resolution full-disk imagery from the \sdo\ Atmospheric Imaging Assembly \citep[AIA;][]{2012SoPh..275...17L}, with a spatial resolution of $1\farcs5$. 1600~\AA\ images were used to study the structure of the flare-associated ribbons. In order to identify the flare-relevant PILs, we inspected filtergram sequences during the flares at UV and EUV wavelengths, in combination with the HMI $B_z$ magnetic field maps. HMI and AIA data was co-registered and co-aligned using standard IDL routines.

\begin{figure*}[ht]
\centerline{
\centering\includegraphics[width=0.9\textwidth]{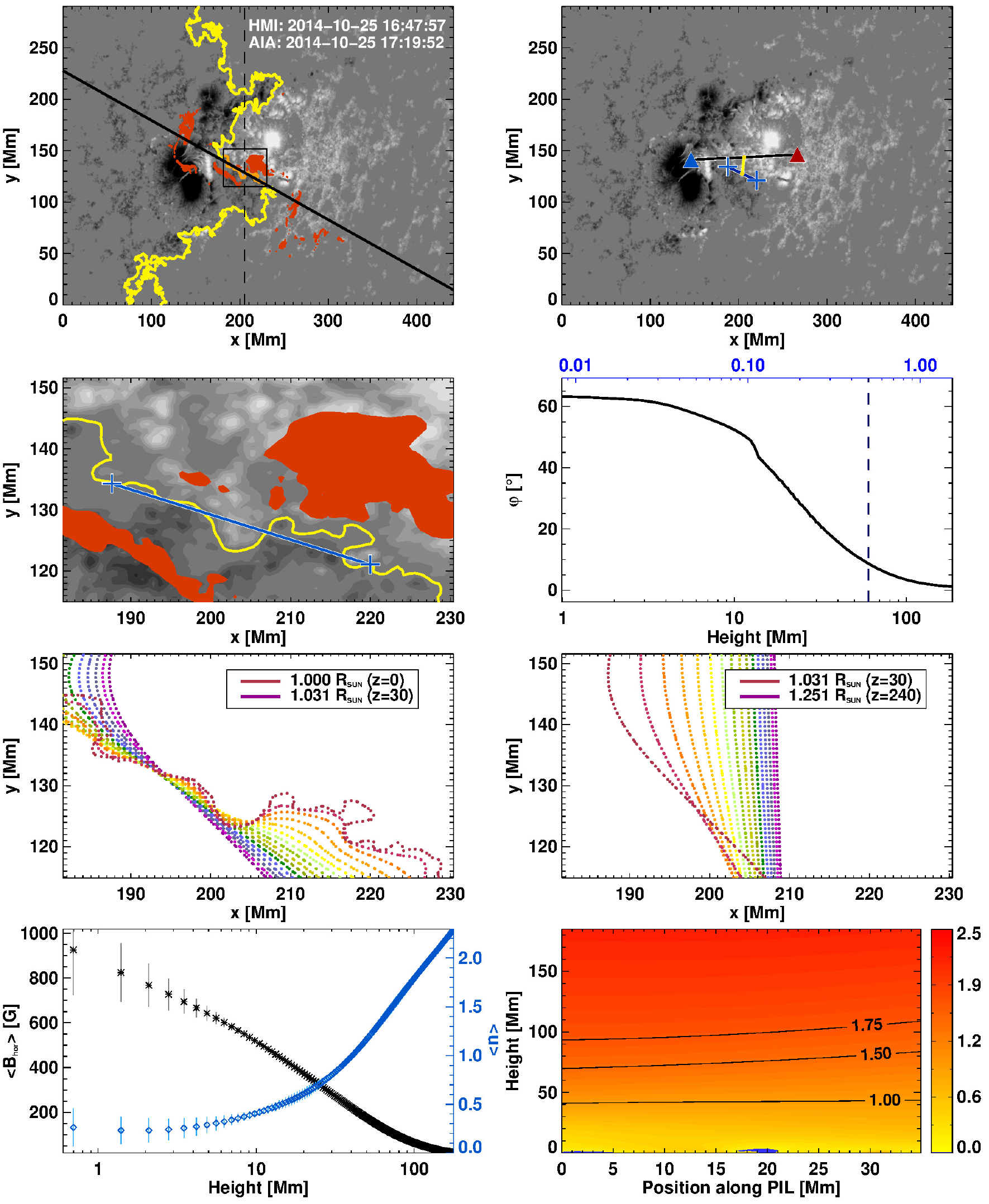}
\put(-467,548){\large\sf(a)}
\put(-467,375){\large\sf(c)}
\put(-467,245){\large\sf(e)}
\put(-467,120){\large\sf(g)}
\put(-230,548){\large\sf(b)}
\put(-230,375){\large\sf(d)}
\put(-230,245){\large\sf(f)}
\put(-230,120){\large\sf(h)}
}
\caption{Analyzed parameters for event no.~15. (a) Linear fit (black line) to the flare-relevant PIL (orange curve). The flare-relevant PIL is determined as a segment of an extended (main) PIL (yellow curve), based on the visibility of flare ribbons observed in AIA~1600~\AA\ (red filled contours). (b) Distance between the flux-weighted AR center and flare site (``flare distance'' $d_{\mathrm{FC}}$; yellow line). Blue/red triangles mark the flux-weighted centers of negative/positive  polarity. Plus signs delimitate the extent of the linear approximation to the flare-relevant PIL (straight blue line connecting plus signs). (c) Linear approximation (blue line connecting plus signs) of the flare-relevant PIL (yellow curve), representing the footprint of a vertical plane used to calculate the mean horizontal magnetic field strength, $\langle B_{hor} \rangle$, and decay index, $n$. The field-of-view is the same as within the black box in (a). (d) PIL orientation, $\varphi$, with respect to the solar North-South direction as a function of height. Labels at the top axis indicate the corresponding normalized height ($h/d_{PC}$). The vertical dashed line indicates the apex height of the underlying dipole field at $h=d_{\mathrm{PC}}/2$. (e) Shape of flare-relevant PIL on 16 selected equidistant height levels in the range 1.0--1.0~$R_{\odot}$ and (f) 1.03--1.25~$R_{\odot}$. (g)  $\langle B_{hor} \rangle$ (black stars) and $\langle n \rangle$ (blue diamonds) as a function of height. Error bars indicate the corresponding standard deviation. (h) Spatial distribution of the decay index above and along of the flare-relevant PIL. Blue areas represent negative values of $n$.}
\label{fig:Overview}
\end{figure*}
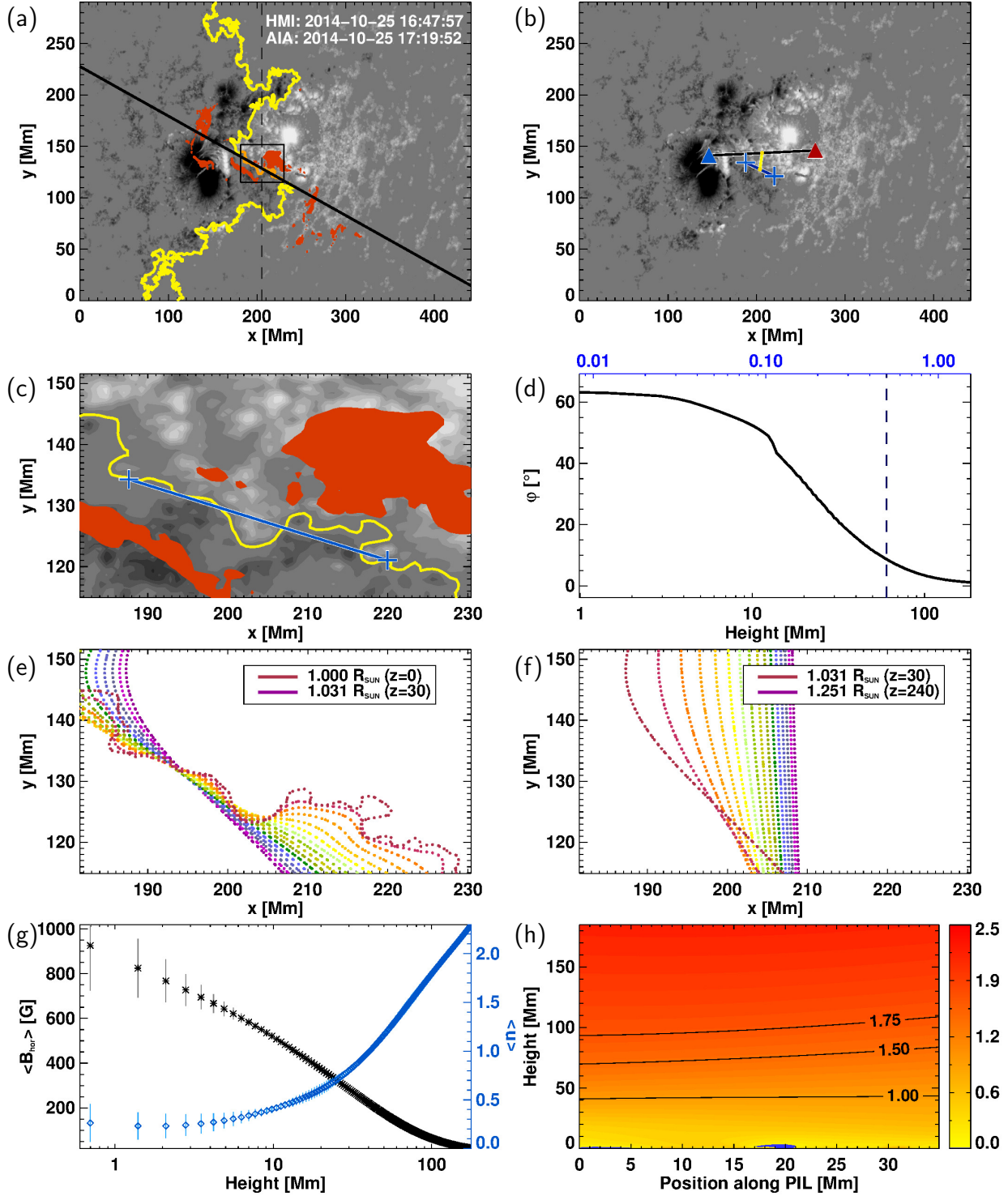

\section{Analysis}
\label{S-Analysis}   

The physical and geometrical parameters described in the following are illustrated using the confined X1.0 flare that occurred on 25 October 2014 
and are visualized in Fig.~\ref{fig:Overview}. The analyzed parameters for all events are listed accordingly in Table~\ref{tab:flare_list}. 

\subsection{Photospheric magnetic field structure\\ and flare location}
\label{SS-DistanceFromPolarityCenter}

To assess the photospheric magnetic field structure of the studied ARs, we first determined the flux-weighted centers of positive and negative magnetic polarity. Here, we considered all pixels that hosted a magnetic field strength $\ge10\%$ of the peak value of the respective polarity (triangles in Fig.~\ref{fig:Overview}b). We calculated the distance between the polarity centers ($d_{\mathrm{PC}}$, black line connecting the triangles in Fig.~\ref{fig:Overview}b). Then, we defined the flux-weighted AR center as the position lying halfway on this connecting line. 

To assess the relative position of the flare site within the host AR, we calculated the ``flare distance'' ($d_{\mathrm{FC}}$; yellow line in Fig.~\ref{fig:Overview}b). $d_{\mathrm{FC}}$ was defined as the distance between the flare site and the flux-weighted AR center. In order to determine the flare site we proceeded as follows. We used flare ribbons observed around the flare peak time in AIA 1600~\AA\ images (red filled contours in Fig.~\ref{fig:Overview}a and \ref{fig:Overview}c), in combination with the $B_z=0$ G contour in the pre-flare HMI map (yellow curve in Fig.~\ref{fig:Overview}a and \ref{fig:Overview}c), to localize the flare-relevant part along the extended PIL within the AR (orange/yellow curve in Fig.~\ref{fig:Overview}a/c). For further analysis, we use a linear approximation of this flare-relevant segment of the PIL (blue line connecting blue crosses on top of the yellow curve in Fig.~\ref{fig:Overview}c). The flare site was then defined as the halfway point along this straight line (position where the yellow and blue lines intersect in Fig.~\ref{fig:Overview}b).

In order to estimate error bounds for $d_{\mathrm{PC}}$ and $d_{\mathrm{FC}}$, we repeated the above described procedure     using the thresholds 5\% and 15\% for the computation of the individual flux-weighted polarity centers.

\subsection{Orientation of the flare-relevant PIL \\ as a function of height}
\label{SS-PIL_InclinationTrend}

In order to study the orientation of the magnetic field in the corona, we traced the flare-relevant PIL as a function of height in our potential magnetic field models. Therefore, we first tracked the flare-relevant PIL at a photospheric level, i.e., at a grid height $z=0$ in our 3D model volume (yellow line in Fig.~\ref{fig:Overview}a). The corresponding PIL at subsequently larger grid heights  ($z=1,\dots,n_z$) was detected using an automated algorithm. At each grid height $z>0$, we searched for the relevant PIL within an area that horizontally extended $\pm 2$~Mm around the PIL that was tracked previously at the next lower grid level (at $z-1$). For example, to find the flare-relevant PIL at model grid height $z=2$, we analyzed B$_z$ at $z=2$ within a $\pm2$~Mm window around the horizontal coordinates of the PIL tracked at $z=1$.

Hereafter, we manually defined a sub-field (black outline in Fig.~\ref{fig:Overview}a), sufficiently large to cover the flare-relevant part of the automatically tracked PIL at each height in the model volume (see Fig.~\ref{fig:Overview}e and \ref{fig:Overview}f for its shape at different heights). Then, we applied a linear fit (straight line in Fig.~\ref{fig:Overview}a) to the flare-relevant PIL at each height level (orange contour in Fig.~\ref{fig:Overview}a). This allowed us to estimate the inclination, $\varphi$, of the PIL with respect to the North-South direction (vertical dashed line in Fig.~\ref{fig:Overview}a) at each height in our model volume (Fig.~\ref{fig:Overview}d). Note that we measured $\varphi$ positively in counter-clockwise direction from solar North. 

Our goal was to quantify how fast the flare-involved magnetic field transits to a configuration aligned with the underlying photospheric magnetic dipole. Naturally, our event sample contains ARs of different size (i.e., of varying $d_{\mathrm{PC}}$). Therefore, we chose to analyze the change of $\varphi$ as a function of physical height, $h$, normalized to the size of the underlying magnetic dipole (see top $x-$ axis in Fig.~\ref{fig:Overview}d). A value $h/d_{\mathrm{PC}}=0.5$ then refers to the top part of the field spanned by the underlying magnetic dipole, i.e., to its apex (vertical dashed line in Fig.~\ref{fig:Overview}d). In that way, we are able to compare the deduced parameters for differently sized ARs with each other in an objective way. In order to deduce the rotation of the flare-relevant PIL over a vertical length scale characteristic for the underlying dipole, we computed $\Delta\varphi=|\varphi(0)-\varphi(h/d_{\mathrm{PC}}=0.5)|$. Here, $\varphi(0)$ denotes the orientation of the flare-relevant PIL at a photospheric level and $\varphi(h/d_{\mathrm{PC}}=0.5)$ is the orientation at the apex of the assumed symmetric dipole field. 

Two main groups are separable in our event sample. One group (62\% of the analyzed events) is comprised of cases for which the flare-relevant PIL was traceable until large heights in the magnetic field models (for instance event no.\ 15; see Fig.~\ref{fig:Overview}d--\ref{fig:Overview}f). For these cases, $\Delta\varphi$ is listed in the last column of Table~\ref{tab:flare_list}. For the remaining cases in our event sample, the flare-relevant PIL could not be traced until a height corresponding to the apex of the underlying dipole field. In other words, the flare-relevant PIL ``disappeared'' already at low heights. Most of these events were found associated to underlying photospheric magnetic field configurations involving parasitic polarities and isolated small-scale magnetic field structures within the ARs. As a result, $\Delta\varphi$ could not be evaluated for these ARs which is denoted by '--' in the last column of Table~\ref{tab:flare_list}.

\subsection{Decay index and flux ratio}
\label{SS-FluxRatioDecayIndex}

In order to characterize the constraining magnetic field in the corona above the flare-relevant PIL, we analyzed the horizontal magnetic field in more detail. We used the linear approximation to the otherwise irregularly shaped flare-relevant PIL (see Sect.~\ref{SS-DistanceFromPolarityCenter} for details) as an input  and assumed this line to represent the footprint of a vertical plane within our magnetic field models. After interpolation of the magnetic field into this plane, we were able to study its strength as a function of height. The black stars in Fig.~\ref{fig:Overview}g give the mean value of the horizontal magnetic field, $\langle B_{hor}\rangle$, as a function of height, taking into account all of the values along the horizontal length of the plane. The error bars mark the corresponding standard deviation.

Based on the horizontal field, we calculated the decay index, $n$, in the vertical plane (Fig.~\ref{fig:Overview}h) and derived the mean decay index, $\langle n \rangle$, along the horizontal length of the plane as a function of height $h$ (blue diamonds in Fig.~\ref{fig:Overview}g; the error bars indicate the 1$\sigma$ standard deviation).  The $\langle n \rangle$ vs.\ $h$ profiles allowed us to determine the critical height for torus instability $h_{\mathrm{crit}}=h(\langle n\rangle_{\mathrm{crit}}=1.5)$. 
The corresponding uncertainties are given by the heights at which $h(\langle n\rangle_{\mathrm{crit}}\pm 1\sigma)$ reaches a value of 1.5.

We note here that the $\langle n \rangle$ vs.\ $h$ profiles in our event sample are almost equally distributed between monotonically increasing (54\%) and saddle-like (46\%) profiles. This is significantly different from the findings of \cite{Wang2017}, who reported these fractions as 85\% and 15\%, respectively, based on an event sample largely comprised of smaller flares (75\% of their analyzed events were $<$M5.0).

We also calculated the horizontal magnetic flux through the vertical plane in the two height regimes $1.0~R_{\odot}\leq h \leq 1.1~R_{\odot}$ ($F_{\rm low}$) and $1.1~R_{\odot}\leq h \leq 1.26~R_{\odot}$ ($F_{\rm high}$), and computed the flux ratio as $F_{\rm low}$/$F_{\rm high}$. The flux ratio has been introduced in \cite{2007ApJ...665.1428W}, with the difference that they estimated $F_{\rm high}$ in the height regime $1.1~R_{\odot}\leq h \leq 1.50~R_{\odot}$. We assessed the effect of this difference on the resulting flux ratio, in that we employed our potential field models until $h=1.5~R_{\odot}$ for a number of events and repeated the flux ratio computation. We found differences in the obtained values of $\approx$10\% at most. The value of the flux ratio also depends on the particular orientation of the vertical plane used to carry out the analysis, i.e., on the direction of the resulting linear fit to the underlying flare-relevant PIL. By comparison of the results for slightly different inclined PILs (all of them resembling the orientation of the underlying flare-relevant PIL to a good degree), the associated uncertainty of the flux ratio is assumed to be $\sim$10\%.

\begin{figure}[t]
\centering\includegraphics[width=\columnwidth]{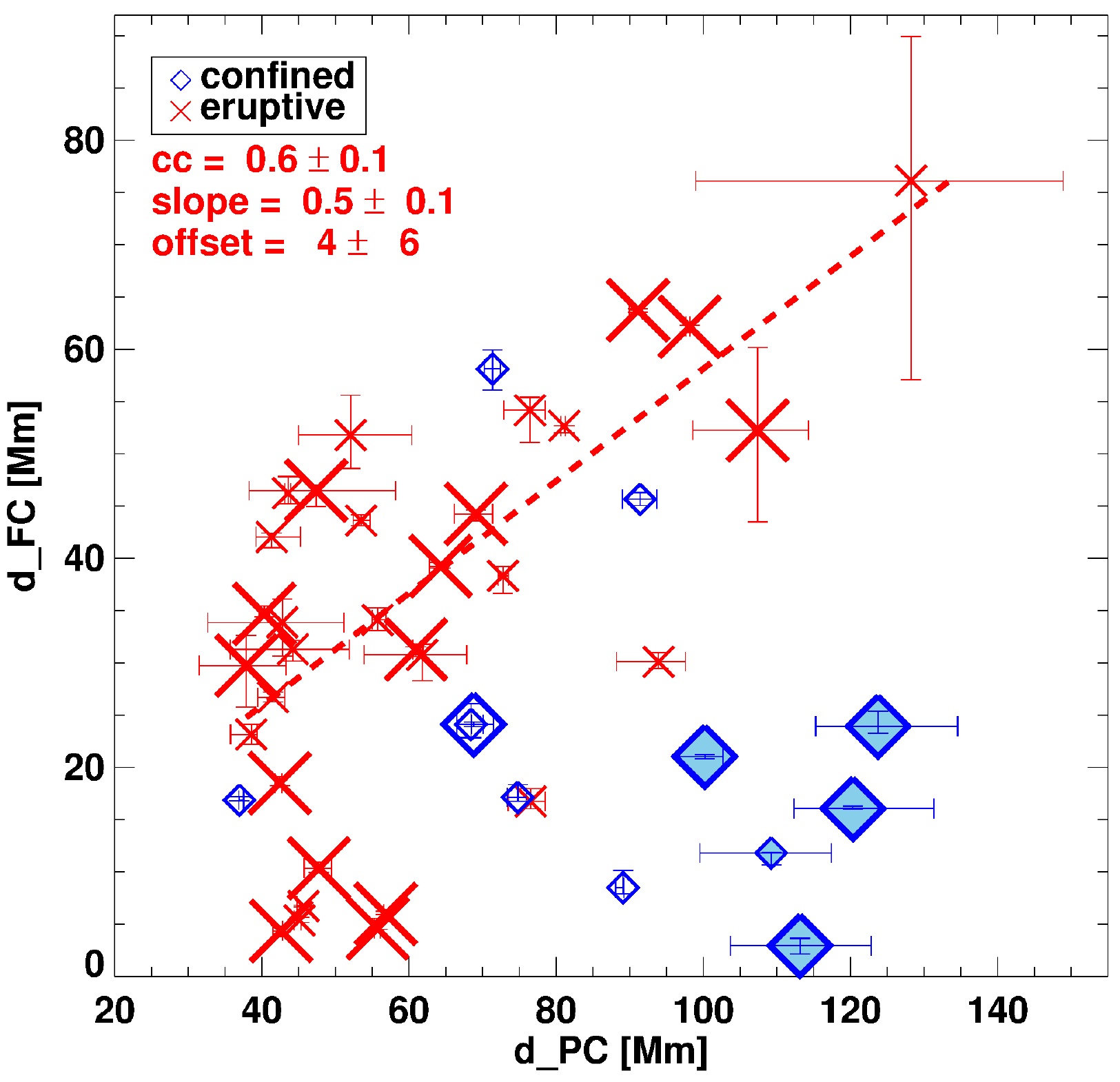}
\put(-246,220){\large\sf(a)}\\
\centering\includegraphics[width=\columnwidth]{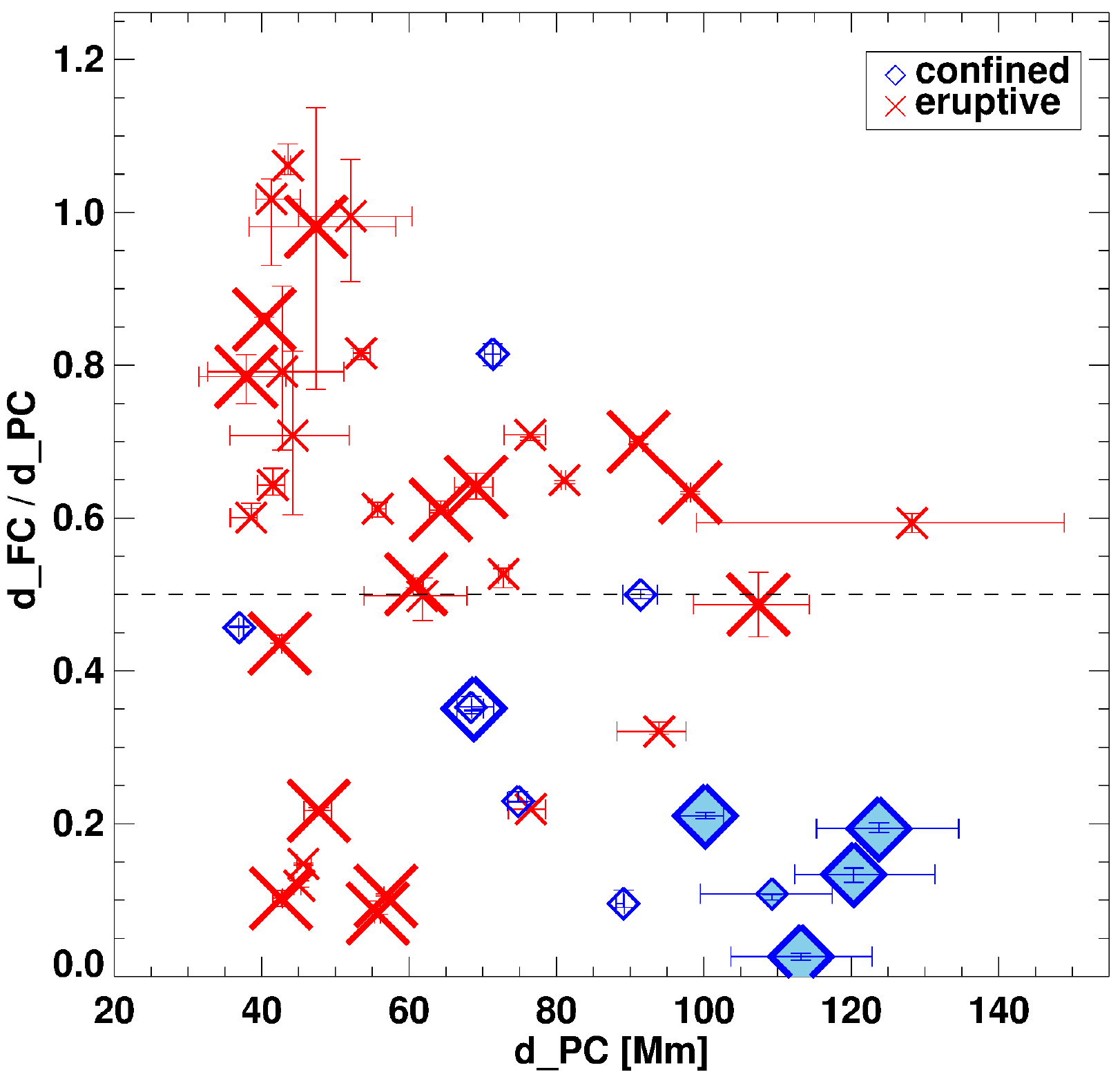}
\put(-246,220){\large\sf(b)}
\caption{Location of all analyzed flares within their host AR. (a) Flare distance ($d_{\mathrm{FC}}$) versus the distance between the centers of opposite magnetic polarity ($d_{\mathrm{PC}}$). Blue diamonds/red stars correspond to confined/eruptive flares. The size of the plot symbols indicates the size of a flare, with smaller/larger symbols indicating M-/X-flares. Filled plot symbols mark the five confined flares that originated from NOAA~12192.
The red solid line represents a linear fit to all eruptive events with the correlation coefficient, $cc_{\rm erup}$, listed. 
(b) Normalized flare distance, $d_{\mathrm{FC}}/d_{\mathrm{PC}}$, vs.\ $d_{\mathrm{PC}}$. The dashed line delimits the area of influence of the underlying bipolar AR magnetic field, if it is approximated by a symmetric magnetic dipole field.}
\label{fig:location}
\end{figure}

\section{Results}
\label{S-Results}

\begin{figure*}[t]
\centering\includegraphics[width=0.85\columnwidth]{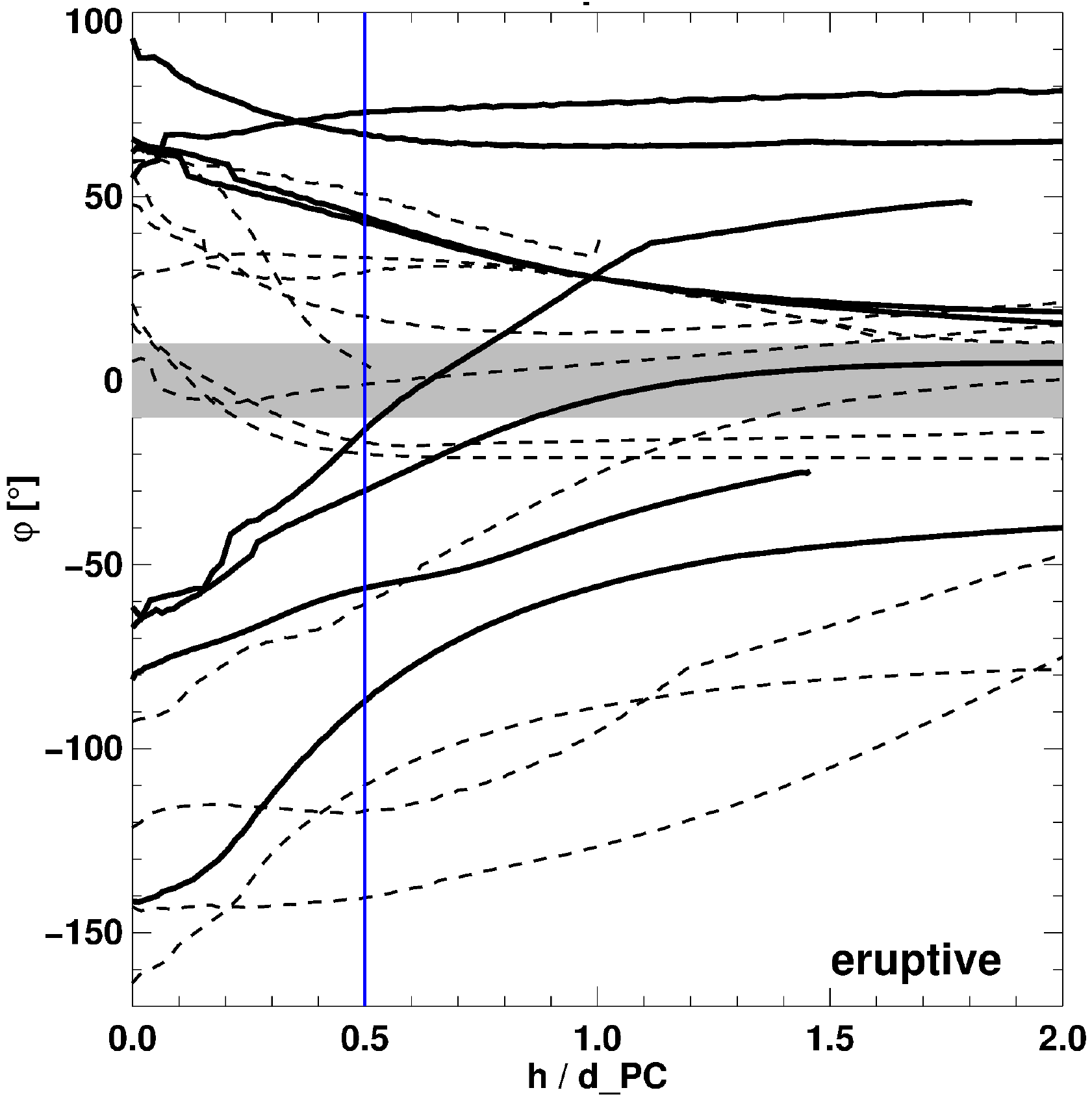}
\centering\includegraphics[width=0.85\columnwidth]{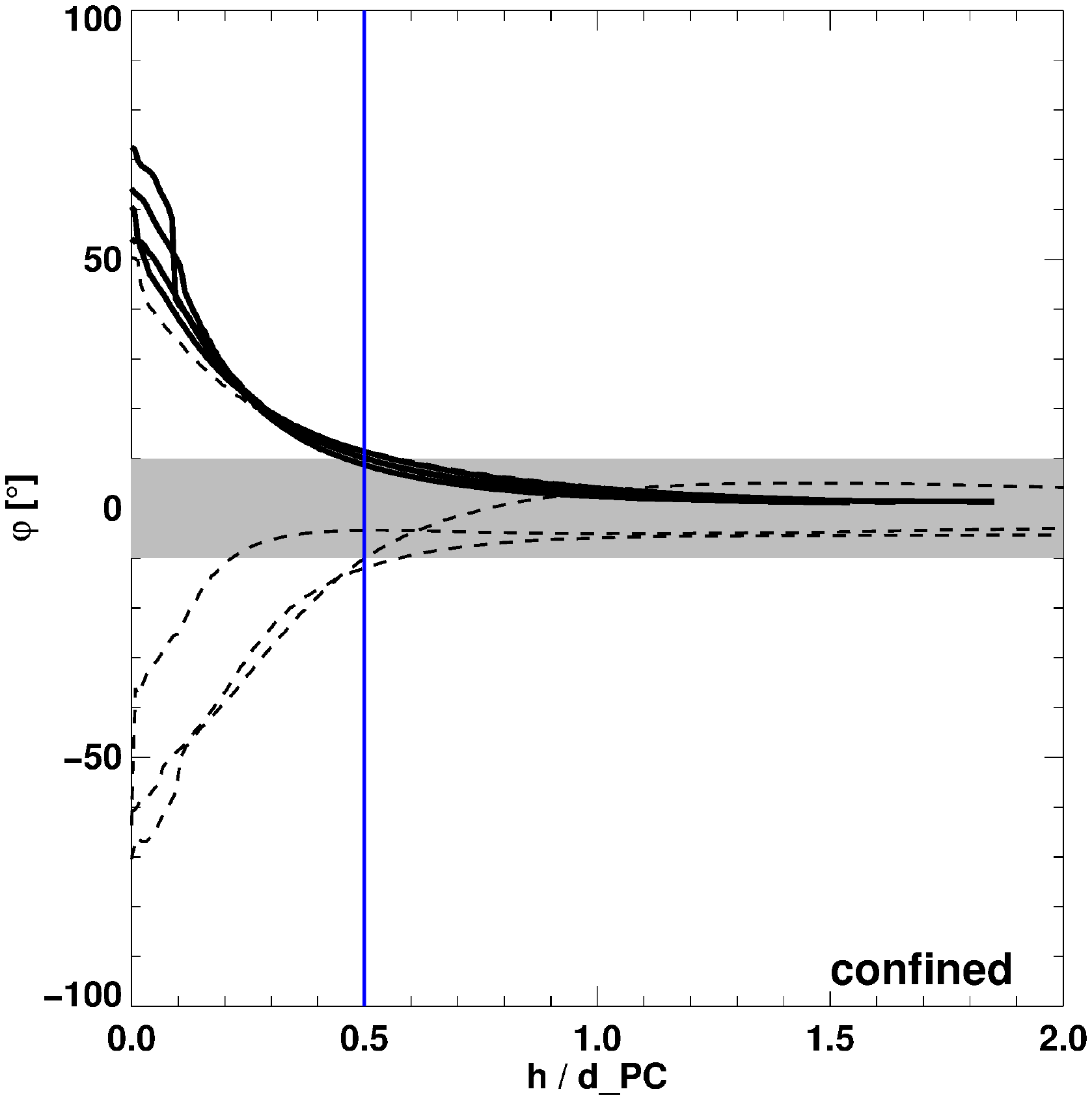}
\put(-420,195){\large\sf(a)}
\put(-210,195){\large\sf(b)}\\
\centering\includegraphics[width=0.85\columnwidth]{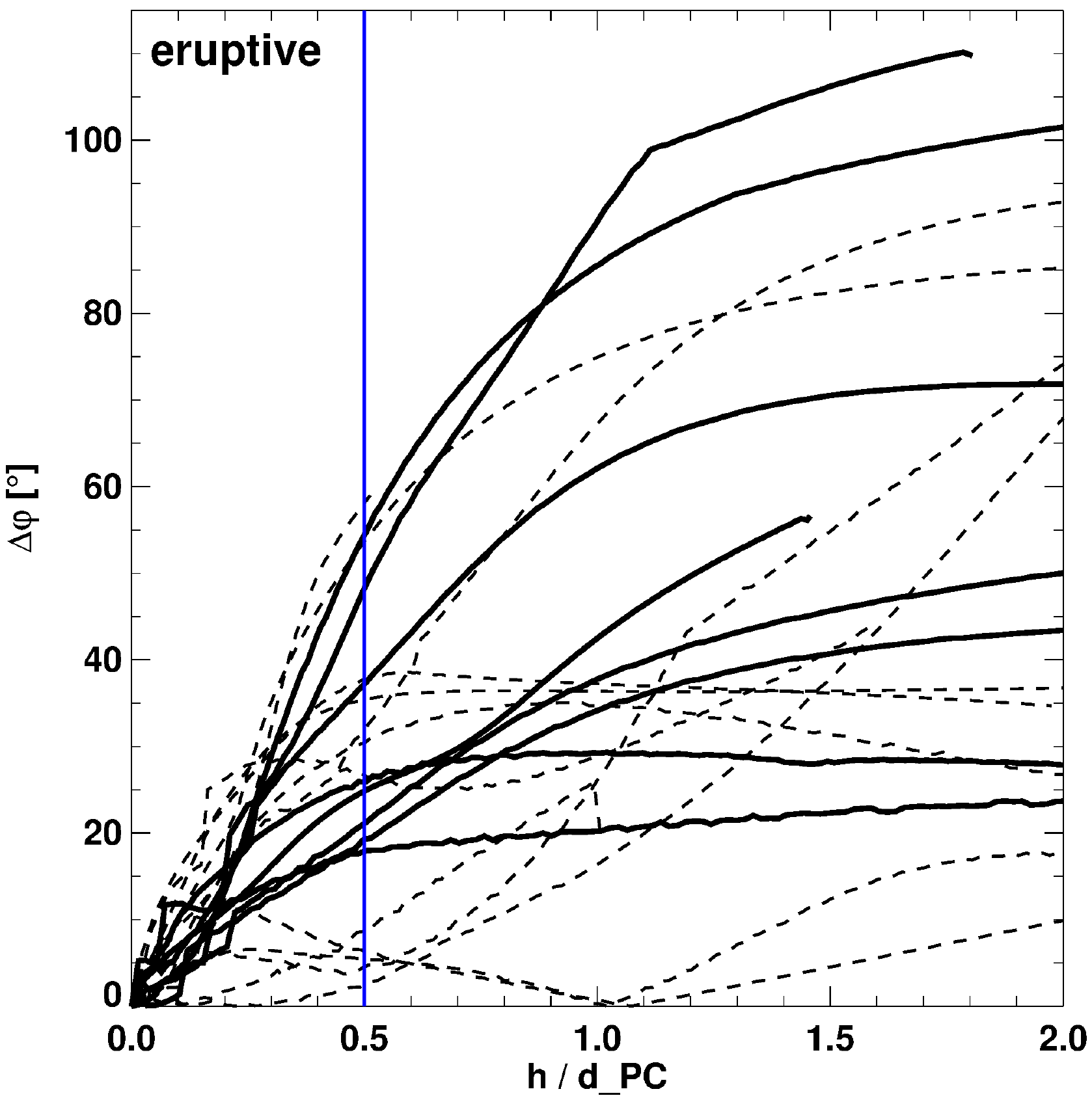}
\centering\includegraphics[width=0.85\columnwidth]{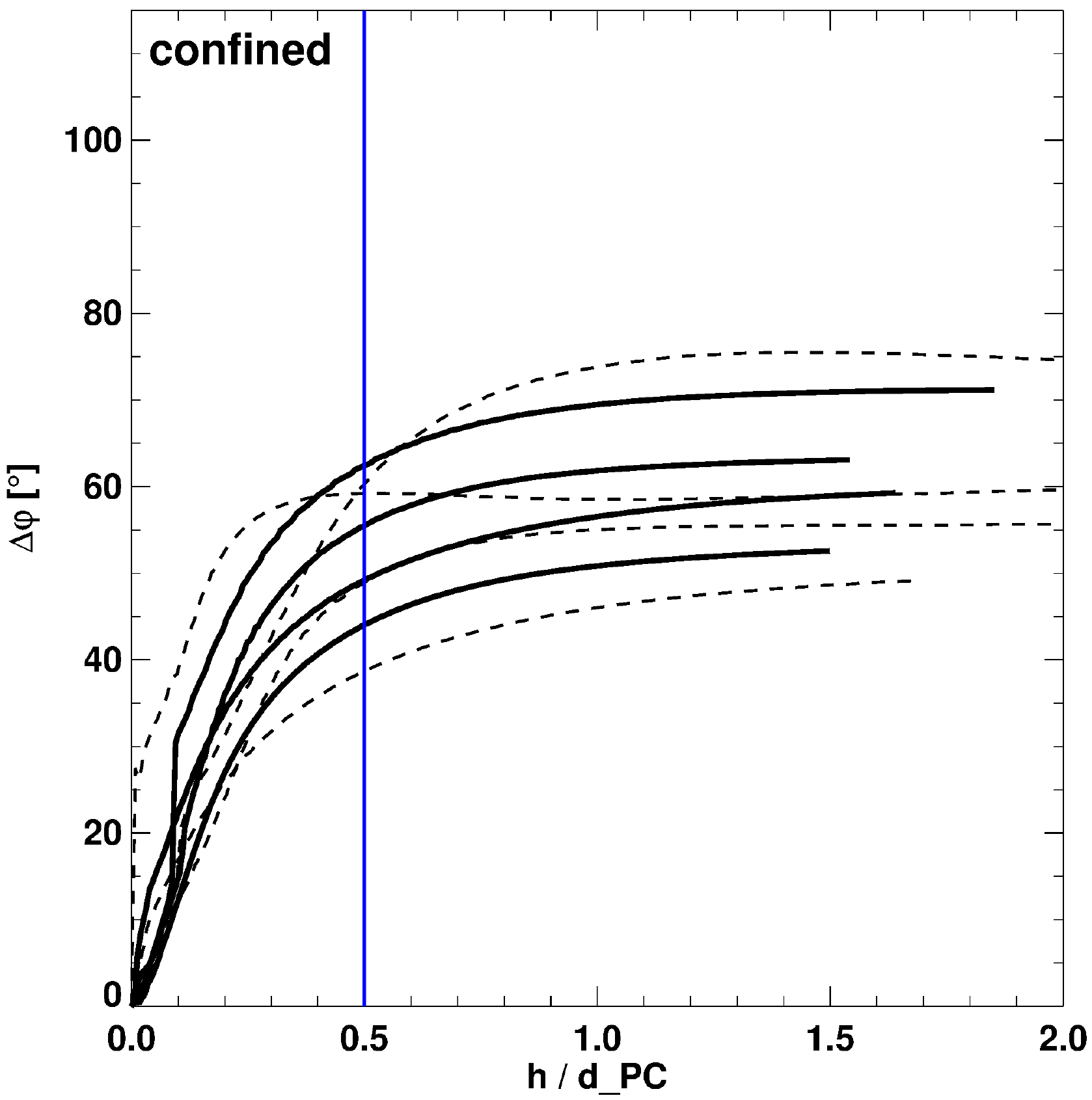}
\put(-420,195){\large\sf(c)}
\put(-210,195){\large\sf(d)}
\caption{Orientation $\varphi$ of flare-relevant PIL vs.\ normalized height $h/d_{\mathrm{PC}}$ for (a) eruptive and (b) confined events. The relative change, $\Delta\varphi=|\varphi(h)-\varphi(0)|$ is shown in (c) and (d), respectively. Results are displayed only for cases for which the flare-relevant PIL was traceable to heights $h/d_{\mathrm{PC}}\ge0.5$. The gray shaded area in (a) and (b) indicates the North-South direction $\pm10^\circ$. The blue vertical line indicates the top of the underlying magnetic dipole field. The dashed and solid lines represent M- and X-flares, respectively.}
\label{fig:INC}
\end{figure*}

In Fig.~\ref{fig:location}a, the flare distance ($d_{\mathrm{FC}}$) is shown in respect to the size of the AR ($d_{\mathrm{PC}}$). Notably, eruptive flares appear to occur in ARs of all sizes, from compact ($d_{\mathrm{PC}}\lesssim 60$~Mm) to extended ARs ($d_{\mathrm{PC}}$ up to $\gtrsim100$~Mm). Confined flares, on the other hand, are found to predominantly originate from extended underlying bipolar field configurations ($d_{\mathrm{PC}}\gtrsim 60$~Mm). 

Eruptive flares are found at a large range of distances ($5\lesssim d_{\mathrm{FC}}\lesssim80$~Mm). In contrast, confined flares originate mainly from close to the flux-weighted magnetic center of the host AR ($d_{\mathrm{FC}}\lesssim25$~Mm). This is not the case in only two confined events: SOL20110730T02:09M9.3 (event no.\ 21) and SOL20120510T04:18M5.7 (event no.\ 29), which can be explained based on the fact that the flares originated from a ``spot-satellite'' region \citep[cf.\ Fig.~6 of][]{2017ApJ...834...56T}. That means that they originated from a minor (``satellite'') bipolar magnetic flux concentration located in the periphery of an AR. This naturally results in a larger value of $d_{\mathrm{FC}}$, if measured with respect to the flux-weighted center of the main AR. 

For the eruptive events we find a clear dependency between $d_{\mathrm{PC}}$ and $d_{\mathrm{FC}}$, with a correlation of $cc\simeq0.6\pm0.1$, independent of the magnitude of the flares (compare size of plot symbols, where smaller/larger symbols denote M-/X-flares). The trend of the flare distance to increase with the size of the host AR might not be a real trend, however, since a larger $d_{\mathrm{FC}}$ can be expected for a more extended host AR (with a larger $d_{\mathrm{PC}}$).  

We fitted a linear regression to the values found for eruptive events in the form $y=a+b\cdot x$, where we proceeded in the following way. We randomized the $x-$ and $y-$ values around the respective values of $d_{\mathrm{PC}}$ and $d_{\mathrm{FC}}$ within their uncertainty ranges and fitted the result using the IDL function {\sc LADFIT}.
Based on $n=10.000$ realizations of this scheme, we determined the mean offset $a=\langle[a_1, \dots, a_n]\rangle$ and mean slope $b=\langle[b_1, \dots, b_n]\rangle$, together with their corresponding uncertainties. For the eruptive events we find $a=0.5\pm0.1$ and $b=4\pm6$. The same approach as outlined here was used to describe the dependencies of the analyzed parameters throughout this section. 

In order to account for the extent of the underlying magnetic dipole field, we show the normalized flare distance $d_{\mathrm{FC}}/d_{\mathrm{PC}}$ in Fig.~\ref{fig:location}b, which also allows us to compare the flare location within an AR for different events. A value $d_{\mathrm{FC}}/d_{\mathrm{PC}}\leq 0.5$ then refers to a location underneath the magnetic field connecting the opposite magnetic polarity centers of the AR.
In case of magnetically compact ARs, this may imply a strong influence of the overlying dipole field. This complies with the fact that we find all but one confined flare at normalized flare distances $d_{\mathrm{FC}}/d_{\mathrm{PC}}\leq0.5$ in Fig.~\ref{fig:location}b. 

In contrast, a value $d_{\mathrm{FC}}/d_{\mathrm{PC}}>0.5$ refers to a location in the periphery of the AR, where the overlying magnetic field should be weaker. And indeed, a large number of eruptive flares are found at such normalized flare distances (red crosses above horizontal dashed line in Fig.~\ref{fig:location}b). Note also that a fraction of the eruptive events originates from close to the magnetic center of their host AR  (red crosses below horizontal dashed line in Fig.~\ref{fig:location}b). This is mostly the case for compact ARs ($d_{\mathrm{PC}}\leq 60$~Mm), however. This is in contrast to confined flares, which tend to predominantly occur close to the flux-weighted magnetic AR center ($d_{\mathrm{FC}}/d_{\mathrm{PC}}\leq0.5$) of  extended ARs ($d_{\mathrm{PC}}\gtrsim 60$~Mm).

Only the confined events no.\ 29 and 44 do not follow this trends ($d_{\mathrm{PC}}\gtrsim60$~Mm but $ d_{\mathrm{FC}}/d_{\mathrm{PC}}\gtrsim0.5$), as well as event no.\ 21 ($d_{\mathrm{PC}}\lesssim60$~Mm and $ d_{\mathrm{FC}}/d_{\mathrm{PC}}<0.5$). As explained above, the spacial organization of the host AR in form of a spot-satellite region explains the deviation from the general tendency. 

Fig.~\ref{fig:INC} shows the orientation of the flare-relevant PIL as function of the normalized height ($h/d_{\mathrm{PC}}$) in our magnetic field models, i.e., with respect to the vertical height of the underlying dipole field. 
We find that in confined flares, the flare-relevant PIL at photospheric levels is inclined by $|\varphi|\gtrsim50^\circ$ (Fig.~\ref{fig:INC}b). In contrast, the flare-relevant PIL in AR's hosting eruptive flares, appear at a large range of possible inclinations (Fig.~\ref{fig:INC}a). Notably, for confined events, we find the PIL to align with the solar North-South direction (indicated by the gray-shaded area in Fig.~\ref{fig:INC}a and \ref{fig:INC}b) rather quickly with height, mostly within a height of $h/d_{\mathrm{PC}}\simeq0.5$, which corresponds to the apex of the surrounding  magnetic dipole field. This is not the case for AR's hosting eruptive flares, where the flare-relevant PIL may still span a large angle with respect the North-South direction ($\varphi\gtrsim10^\circ$ at $h/d_{\mathrm{PC}}=0.5$ in Fig.~\ref{fig:INC}a). In other words, the PIL orientation changes much faster with height above ARs that host confined flares than above those that host eruptive flares. 

\begin{figure}[t]
\centering\includegraphics[width=\columnwidth]{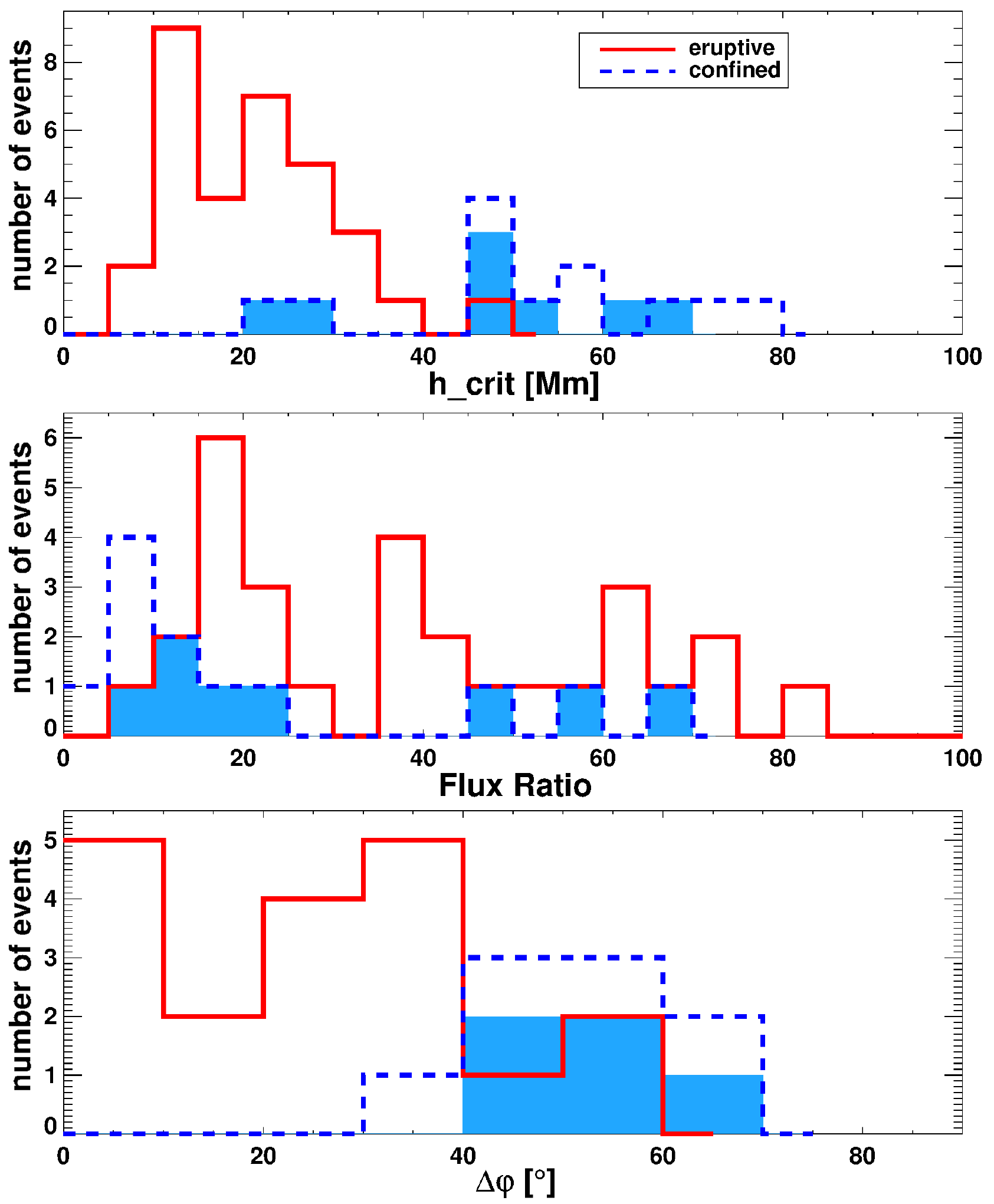}
\put(-30,278){\large\sf(a)}
\put(-30,180){\large\sf(b)}
\put(-30,82){\large\sf(c)}
\caption{Histograms for (a) $h_{\mathrm{crit}}$, (b) flux ratio and (c) $\Delta\varphi$. Red (solid) and blue (dashed) lines represent relative occurrences of eruptive and confined flares, respectively. { The filled histograms show the relative distribution for confined flares if the individual respective values of the five flares that originated from NOAA~12192 are replaced by the respective mean values.}} 
\label{fig:histogram}
\end{figure}

In Fig.~\ref{fig:INC}c and \ref{fig:INC}d we plot the change of the orientation of the flare-relevant PIL with respect to the photospheric level, $\Delta\varphi=|\varphi(h)-\varphi(0)|$, i.e., with respect to the orientation of the underlying dipole field. For confined events, we find most prominent changes ($\Delta\varphi\simeq40^\circ$--$60^\circ$) of the PIL orientation to occur over a height characteristic of the underlying dipole field ($h/d_{\mathrm{PC}}\lesssim0.5$; Fig.~\ref{fig:INC}d). This is not the case for eruptive flares, where the changes range from a few degrees up to about $60^\circ$ (Fig.~\ref{fig:INC}c). Moreover, the orientation of the flare-relevant PIL may still undergo strong changes above the apex of the underlying dipole field (above $h/d_{\mathrm{PC}}=0.5$). In Fig.~\ref{fig:histogram}c, we show the distribution of $\Delta\varphi$, separately for confined and eruptive flares (blue and red histograms, respectively). We find the distributions of the two groups of events as distinct, with the confined events joining the distribution of eruptive events at large values ($\Delta\varphi\gtrsim40$). In particular, we find $\Delta\varphi\lesssim40^\circ$ for most of the ARs that host eruptive events and $\Delta\varphi\gtrsim40^\circ$ for those that host confined events (compare also Fig.~\ref{fig:INC}c and \ref{fig:INC}d). We find mean values of $\langle\Delta\varphi\rangle=26.1\pm16.4^\circ$ and $\langle\Delta\varphi\rangle=53.1\pm8.2^\circ$, respectively. 

Note that five confined flares in our event sample (events no.\ 13--16 and 37) originated from a single AR (NOAA~12192) over four days, and thus exhibit very similar values of the deduced parameters. In order to compensate for this over-representation in our { confined events} statistics, we computed the mean values of $\Delta\varphi$ from the individual values of the five events and show the corresponding occurrences as filled histograms in Fig.~\ref{fig:histogram}c (and in a similar way for $h_{\mathrm{crit}}$ and the flux ratio in Fig.~\ref{fig:histogram}a and b, respectively). When considering this ``reduced'' set of values,  we obtain $\langle\Delta\varphi\rangle=55.5\pm5.5^\circ$.

In Fig.~\ref{fig:histogram}a we show the distribution for $h_{\mathrm{crit}}$ which well separates between AR's that host confined ($h_{\mathrm{crit}}\gtrsim40$~Mm) and eruptive ($h_{\mathrm{crit}}\lesssim40$~Mm) events, with only a small overlap of the populations. We find mean values of $\langle h_{\mathrm{crit}}\rangle=52.7\pm16.7$~Mm and $\langle h_{\mathrm{crit}}\rangle=20.9\pm9.5$~Mm for ARs that hosted confined and eruptive events, respectively. Compensation for the over-represented NOAA~12192 gives $\langle h_{\mathrm{crit}}\rangle=47.4\pm16.0$~Mm. The corresponding histogram of the flux ratio (Fig.~\ref{fig:histogram}b) shows confined and eruptive events as two distinct but overlapping populations, with the confined events residing at the lower range of values (flux ratio $\lesssim25$)
Only a few confined events, no.\ 21 (${\rm flux~ratio}\simeq58$), 29 (${\rm flux~ratio}\simeq49$), and 44  (${\rm flux~ratio}\simeq67$) show larger values. Again, these are the confined flare events that originated from spot-satellite ARs.

In order to test whether or not the distributions of $h_{\mathrm{crit}}$, flux ratio and $\Delta\varphi$ for confined and eruptive events can be modeled to originate from a single, normally distributed population of events, we performed an Anderson-Darling 2-sample test (using the package “kSamples” in R). The $p$-values obtained for the three considered quantities are found not to exceed $0.01$. In other words, we may reject the corresponding null-hypothesis with 95\% confidence and consider the results obtained for confined and eruptive events as to be distinctly different.

\begin{figure}[t]
\centering\includegraphics[width=\columnwidth]{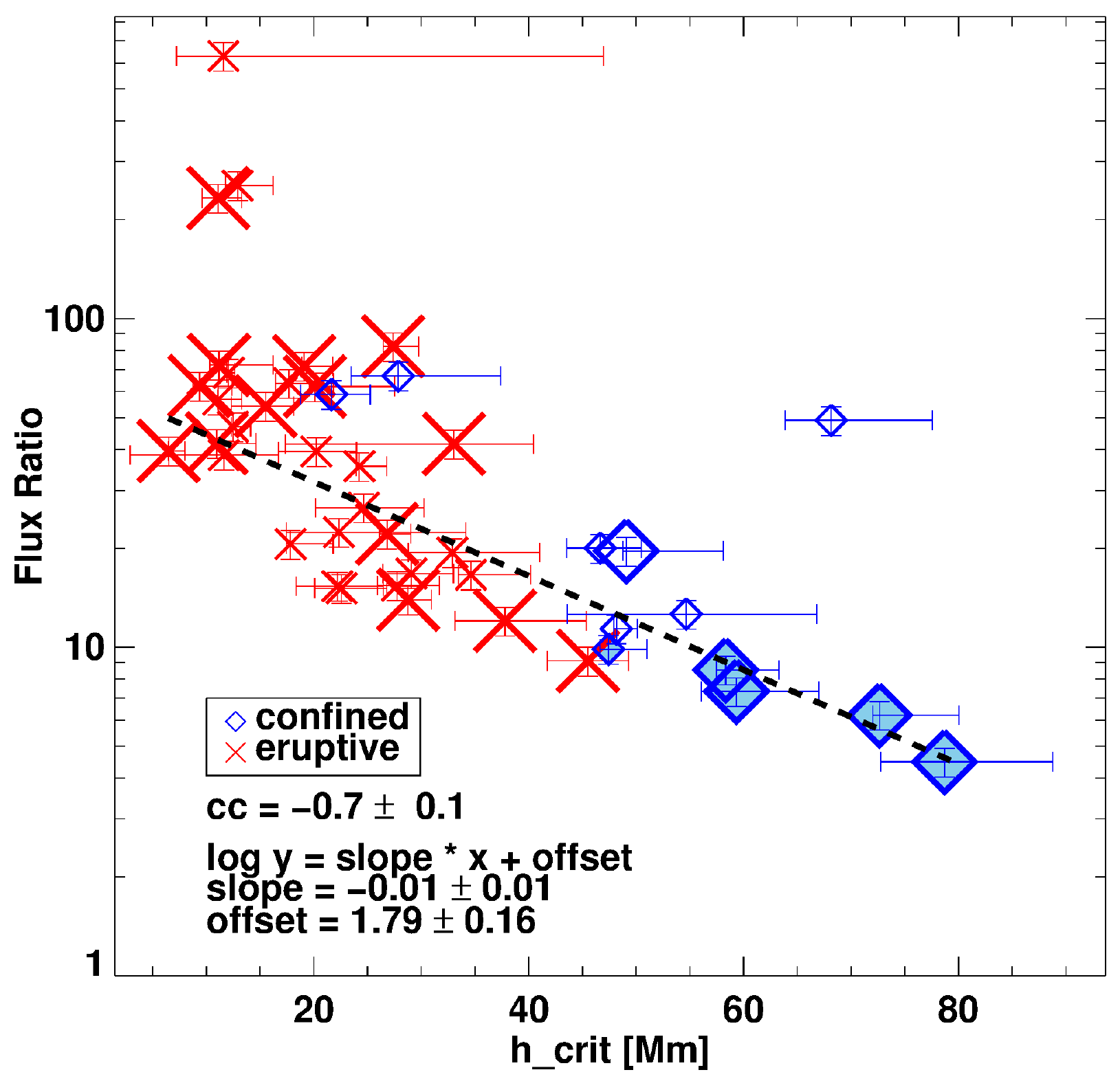}
\put(-242,220){\large\sf(a)}\\
\centering\includegraphics[width=\columnwidth]{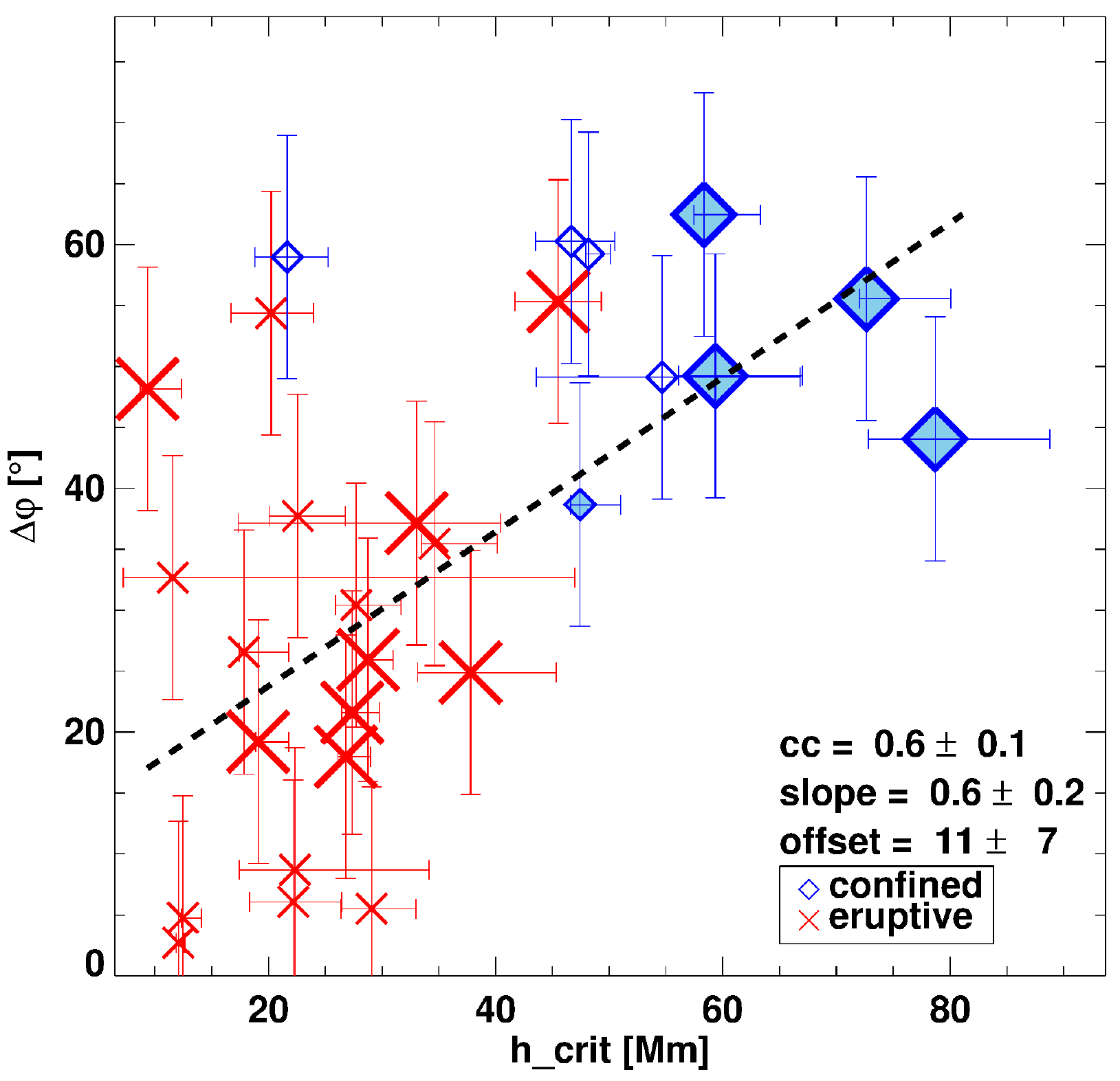}
\put(-242,220){\large\sf(b)}\\
\caption{(a) Flux ratio versus critical height for torus instability, $h_{\mathrm{crit}}$. (b) Change of orientation of the flare-relevant PIL, $\Delta\varphi$, versus $h_{\mathrm{crit}}$. Black solid lines indicate a linear fit to all data points with respective correlation coefficients, $cc$, listed. The corresponding uncertainties were computed using a bootstrapping method. Blue diamonds and red stars correspond to confined and eruptive flares, respectively. The size of the symbols indicates the size of the corresponding flare, with smaller/larger symbols indicating M-flares/X-flares. Filled plot symbols mark the five confined flares that originated from NOAA~12192.
}
\label{fig:CRITvs}
\end{figure}

Last, we check for possible relationships between the different deduced quantities. We find a clear correspondence between $h_{\mathrm{crit}}$ and the flux ratio ($cc\simeq-0.7\pm0.1$; see Fig.~\ref{fig:CRITvs}a). Since the data points well align with the linear fit (black line) we may safely assume that both parameters are robust indicators of the decay of the horizontal magnetic field overlying a flaring AR. However, $h_{\mathrm{crit}}$ appears to be a much better discriminator between AR's that host confined or eruptive flares. Also $h_{\mathrm{crit}}$ and $\Delta\varphi$ are correlated ($cc\simeq0.6\pm0.1$; see Fig.~\ref{fig:CRITvs}b). In that case, however, the correlation appears to result mainly from the apparent clustering of the parameters deduced for confined and eruptive flares in the distinct parts of the scatter plot. While confined events are concentrated in the upper right quadrant ($h_{\mathrm{crit}}\gtrsim40$~Mm, $\Delta\varphi\gtrsim40^\circ$), the parameters for the majority of eruptive cases is found in the lower left quadrant.

After compensation for NOAA~12192 being over-represented, we find $cc{(h_{\mathrm{crit}},{\rm flux~ratio})}\simeq-0.4\pm0.1$ and $cc{(h_{\mathrm{crit}},\Delta\varphi)}\simeq0.2\pm0.1$.

\section{Discussion}
\label{S-Discussion}

We aimed at a better understanding of the magnetic field structure in and above flaring ARs in terms of the type of flare activity---confined or eruptive. We analyzed 44 large flares (\goes\ class M5.0 and larger) that occurred between January 2011 and December 2015, which were observed close to the solar disk center, and for which the flare type could be unambiguously determined. 

Our goal was to deepen our insights regarding the discriminating factors in the flare-associated magnetic field, both at photospheric and coronal levels. For this purpose, we constructed a 3D potential field magnetic field model for each flare AR, based on a pre-flare photospheric {\it SDO}/HMI vector magnetic field map.  Importantly, we put all considerations in context with the characteristic length scale of the bipolar magnetic field configuration of the host AR, which we approximated by a symmetric dipole with a typical length scale equal to $d_{\mathrm{PC}}/2$, both in horizontal and vertical direction (height). Here, $d_{\mathrm{PC}}$ denotes the distance between the flux-weighted centers of opposite magnetic polarities within the AR.

At photospheric levels, we investigated the distance of the flare site to the flux-weighted AR polarity center ($d_{\mathrm{FC}}$), as well as the size of the host AR (given by $d_{\mathrm{PC}}$). At coronal levels, we analyzed quantities characterizing the relative strength of the confining magnetic field (decay index $n$ and flux ratio) and the change of orientation of the flare-relevant PIL as a function of height ($\Delta\varphi$). To our knowledge, $\Delta\varphi$, has not been investigated so far systematically for a large number of flare events. 

Perhaps our most important finding is a distinctly different spatial organization of the magnetic field above ARs that host confined or eruptive flares. In ARs that host confined flares, the orientation of the flare-relevant PIL changes much faster with height towards a direction representative for the underlying dipole magnetic field. The orientation of the PIL changes quickly  from an initial high inclination at photospheric levels to a direction representing the underlying dipole field, within a typical length scale of the underlying bipolar magnetic field ($\Delta\varphi\gtrsim40^\circ$ until a height $h\simeq d_{\mathrm{PC}}/2$; Fig.~\ref{fig:INC}d). This is not the case for ARs that host eruptive flares. There, the transition towards the characteristic orientation of the underlying dipole field involves much larger heights (Fig.~\ref{fig:INC}c). 
$\Delta\varphi$ discriminates to a certain degree between AR's that host confined and eruptive flares, showing two populations in the corresponding histogram that overlap at large values ($40^\circ\lesssim\Delta\varphi$; cf.\ Fig.~\ref{fig:histogram}c). This indicates that the relative orientation of the coronal field at different heights above a flare site may play a role in determining whether a confined or eruptive flare will occur. 

\begin{figure}[t]
\centering\includegraphics[width=\columnwidth]{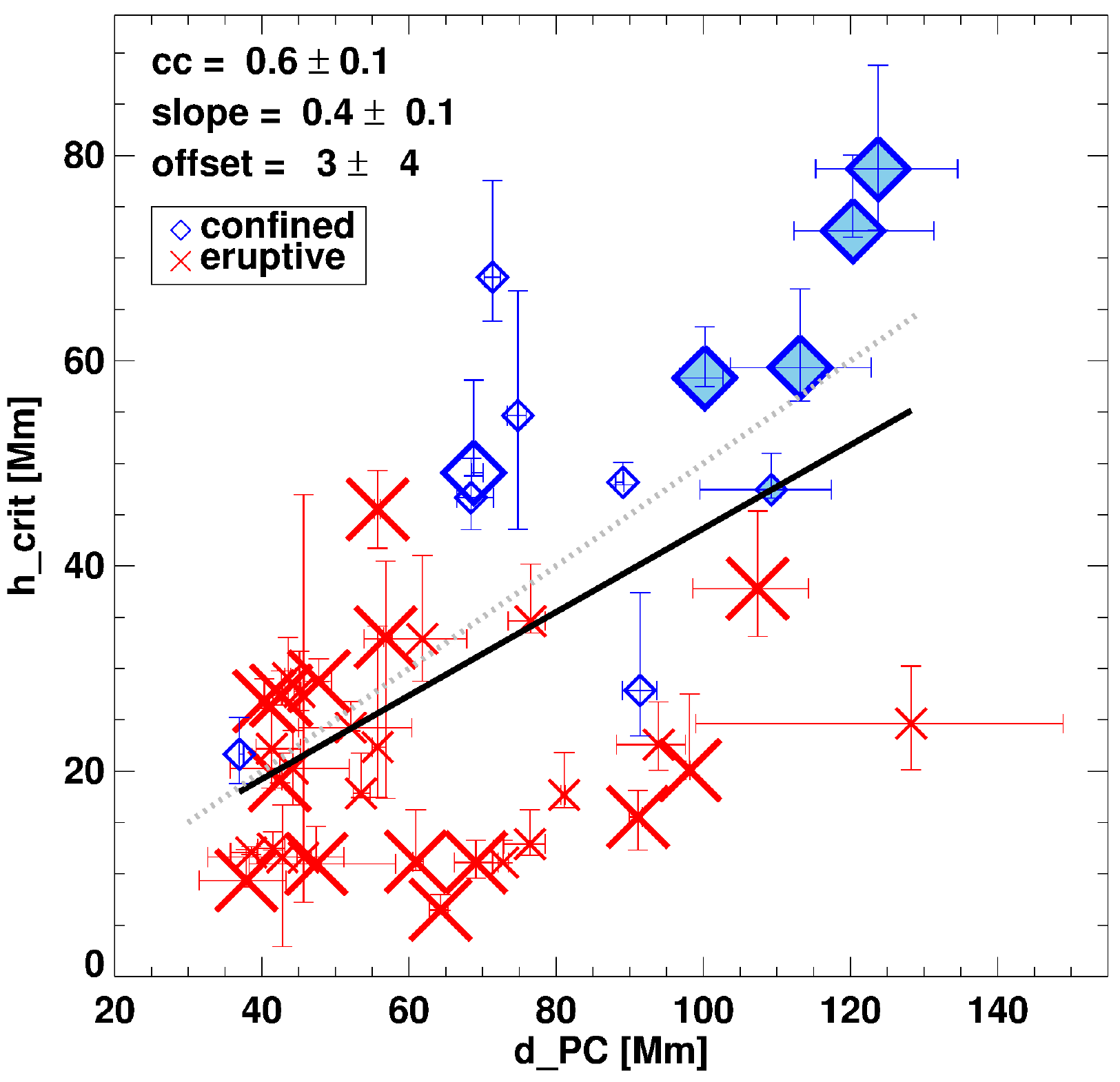}
\caption{Critical height of torus instability, $h_{\mathrm{crit}}$, versus distance between the flux-weighted centers of opposite polarity, $d_{\mathrm{PC}}$. The black solid line indicates a linear fit to all data points. The gray dashed line represents the empirical relation $h_{\mathrm{crit}}=d_{\mathrm{PC}}/2$. Blue diamonds and red stars correspond to confined and eruptive flares, respectively. The size of the symbols indicates the size of the corresponding flare. Smaller/larger symbols indicate M-flares/X-flares. Filled plot symbols mark the five confined flares that originated from NOAA~12192.
}
\label{fig:HCReferences}
\end{figure}

Equally important is the finding that the flare type can be easily understood in terms of the flare site in respect to the size of the dipole magnetic field of the host AR. We found that confined flares mainly originate from extended ARs ($d_{\mathrm{PC}}\gtrsim60$~Mm), while eruptive flares originate from ARs of very different sizes ($40\lesssim d_{\mathrm{PC}}\lesssim140$~Mm; see Fig.~\ref{fig:location}). Also, confined flares primarily originate from locations close to the flux-weighted AR center ($d_{\mathrm{FC}}\lesssim25$~Mm), while eruptive flares may originate from basically everywhere within an AR ($5\lesssim d_{\mathrm{FC}}\lesssim80$~Mm). In order to address the importance of the location of the flare with respect to the underlying bipolar magnetic field, we considered normalized flare distances ($d_{\mathrm{FC}}/d_{\mathrm{PC}}$). This allowed us first, to place the flare site with respect to the possibly weak or strong confinement of the surrounding magnetic field and second, to make the measured distances and extensions of the considered ARs comparable to each other.

The use of normalized distances places the confined events clearly inside of the underlying magnetic dipole field ($d_{\mathrm{FC}}/d_{\mathrm{PC}}<0.5$; Fig.~\ref{fig:location}b). Since the overlying magnetic field in extended ARs ($d_{\mathrm{PC}}\gtrsim60$~Mm) decays slowly with height, a correspondingly strong confinement exits that likely prohibits the escape of an possibly associated CME. In contrast, a large number of eruptive flares originates from the periphery of ARs ($d_{\mathrm{FC}}/d_{\mathrm{PC}}>0.5$) where the overlying field is weaker. Notably, eruptive flares that originate from within the strong confinement of the underlying AR dipole field ($d_{\mathrm{FC}}/d_{\mathrm{PC}}<0.5$) mostly do so in compact ARs ($d_{\mathrm{PC}}\lesssim60$~Mm). Here, we may argue that in compact ARs, the confining magnetic field decays rapidly with height, i.e., the critical height for torus instability $h_{\mathrm{crit}}$, resides at lower coronal heights, allowing for flare-associated CMEs.

Correspondingly, we find clearly lower values of $h_{\mathrm{crit}}$ in ARs that host eruptive flares ($\langle h_{\mathrm{crit}}\rangle\simeq21\pm10$~Mm) than for those which host confined flares ($\langle h_{\mathrm{crit}}\rangle\simeq53\pm17$~Mm), in accordance to earlier studies \citep[e.g.][]{2011ApJ...732...87C, Wang2017}. \cite{Wang2017} suggested an empirical relation of the form $h_{\mathrm{crit}}\simeq 0.5\cdot d_{\mathrm{PC}}$ (cf.\ their Fig.~2 and see gray dotted line in Fig.~\ref{fig:HCReferences}). 
Based on our event sample and analysis, we find $h_{\mathrm{crit}}\simeq 0.4\pm0.1 \cdot d_{\mathrm{PC}}$ (black solid line in Fig.~\ref{fig:HCReferences}), which agrees with the trend found in \cite{Wang2017} within our uncertainty range. As a reason for the slightly different result, we suspect the different flare sizes covered in our study. \cite{Wang2017} analyzed 60 flares down to \goes\ class M1.0, where the majority of events (75\%) were classified as $<$M5.0. In our study we considered exclusively large events, i.e., flares of  \goes\ class M5.0 and larger.

Another important finding in our study is that $h_{\mathrm{crit}}$ discriminates better between AR's that host confined and eruptive flares than the flux ratio, an alternative quantity to measure the decay of the horizontal field (Fig.~\ref{fig:CRITvs}a), which has been introduced by \cite{2007ApJ...665.1428W}. Importantly, $h_{\mathrm{crit}}$ shows two distinct populations, and clearly separates between confined ($h_{\mathrm{crit}}\gtrsim40$~Mm) and eruptive flares ($h_{\mathrm{crit}}\lesssim40$~Mm; Fig.~\ref{fig:histogram}a). The flux ratio on the other hand shows a less clear distinction (compare Fig.~\ref{fig:histogram}b). We suspect the reason in the higher sensitivity of $h_{\mathrm{crit}}$ towards the non-linear decay of the horizontal field with height in the corona. 

In order to substantiate whether or not the studied parameters deduced from ARs hosting confined and eruptive events can be used as discriminating factors at all (i.e., whether they stem from a single normally or non-normally distributed population of events), we performed an Anderson-Darling 2-sample test for $h_{\mathrm{crit}}$, flux ratio, and $\Delta\varphi$. This allows us to state with 95\% confidence that the studied parameters do not fit a normal distribution, i.e., can be considered as to be distinctly different for confined and eruptive events.

\begin{figure}[t]
\centering\includegraphics[width=\columnwidth]{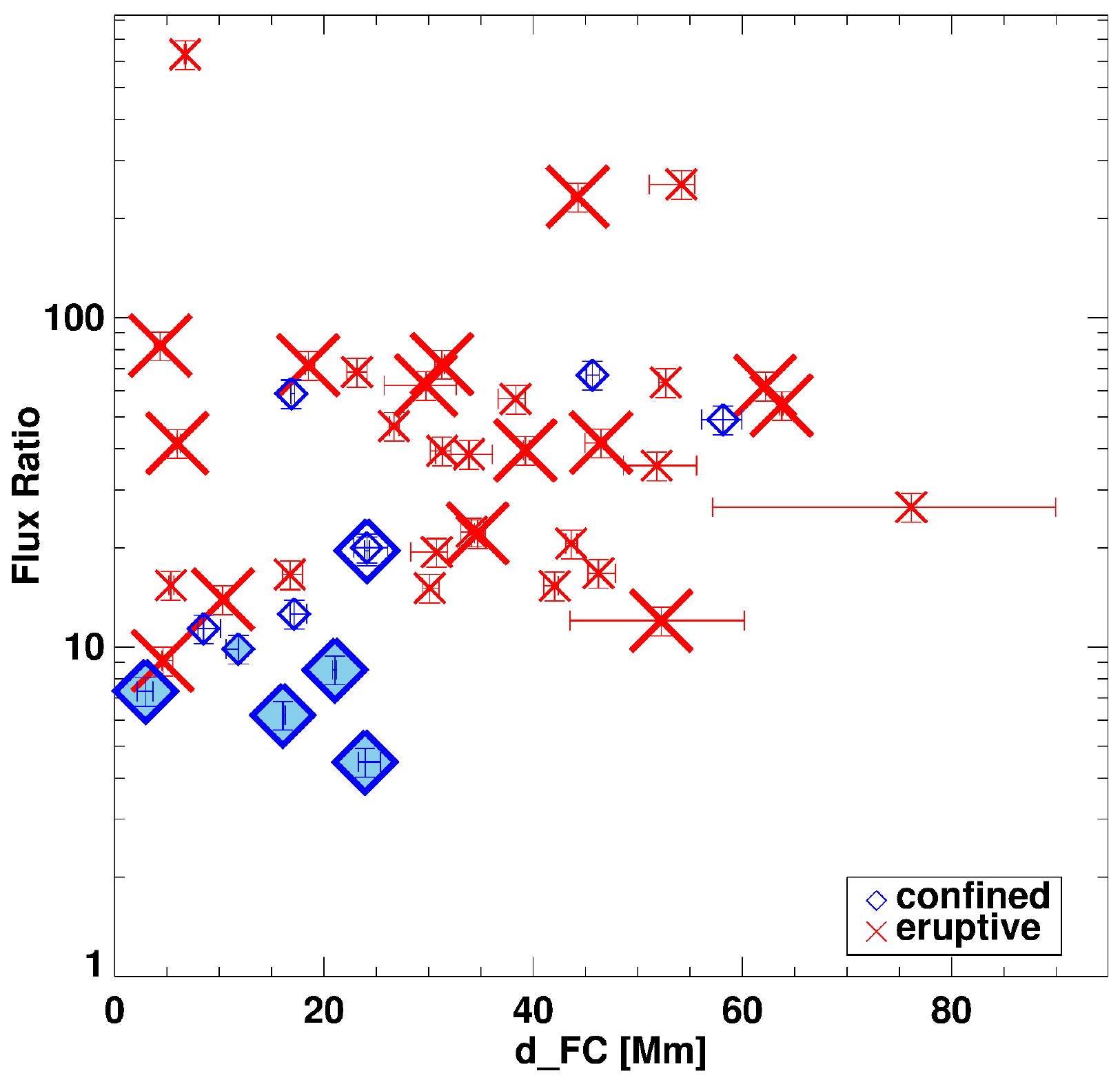}
\caption{Flux ratio as a function of the distance between the flare site and the flux-weighted AR center, $d_{\mathrm{FC}}$. Blue diamonds and red stars correspond to confined and eruptive flares, respectively. The size of the symbols indicates the size of the corresponding flare. Smaller/larger symbols indicate M-flares/X-flares. Filled plot symbols mark the five confined flares that originated from NOAA~12192.
}
\label{fig:FRReferences}
\end{figure}

\cite{2007ApJ...665.1428W} furthermore reported a clear segregation of confined and eruptive events in a flux ratio vs.\ $d_{\mathrm{FC}}$ diagram (cf.\ their Fig.~5). More precisely, they found the confined/eruptive events to reside at lower/higher values of both, $d_{\mathrm{FC}}$ and flux ratio. We do recover this trend only to a certain degree, and with a large overlap of the two populations (see Fig.~\ref{fig:FRReferences}). A reason for the differing results may again be the particular event selection, on which the different studies are based on. While \cite{2007ApJ...665.1428W} analyzed only 8 flares, we considered a much larger number of events ({ 44} in total). Also, while \cite{2007ApJ...665.1428W} studied only major (X-class) flares, we included large flares (M5.0-class and larger). Importantly, the overlap in our diagram is not solely caused by the eruptive M-flares included in our study, i.e., we also find eruptive X-flares at low values of both, $d_{\mathrm{PC}}$ and flux ratio. Consequently, the different finding  cannot be simply attributed to the larger range of flare sizes included in our study.

\section{Summary and Conclusions}
\label{S-Summary}

We analyzed { 44} flares (\goes\ class $\ge$M5.0) that occurred between January 2011 and December 2015. Based on \sdo/HMI vector magnetic field maps, we modeled the 3D potential magnetic field above the flare ARs. We studied the vertical decay and orientation of the flare-relevant 3D magnetic field, in relation to the associated flare type, confined or eruptive. Our most important findings include:

1. The flare type may be well understood in terms of the flare site within an AR, measured by the flare distance from the flux-weighted AR center, $d_{\mathrm{FC}}$. This is especially true if $d_{\mathrm{FC}}$ is interpreted with respect to the characteristic length scale of the underlying photospheric magnetic dipole (approximated by the half distance between the centers of opposite magnetic polarity, $d_{\mathrm{PC}}/2$). If located underneath the confining field spanned by the magnetic dipole (i.e., at a distance $d_{\mathrm{FC}}/d_{\mathrm{PC}}\leq0.5$), flares tend to be eruptive when hosted by a compact ($d_{\mathrm{PC}}\lesssim60$~Mm) AR and confined when launched from an extended AR ($d_{\mathrm{PC}}\gtrsim60$~Mm). Flares that originate from the periphery of the confining bipolar field ($d_{\mathrm{FC}}/d_{\mathrm{PC}}>0.5$) tend to be eruptive, no matter if the AR is compact or not.

2. The critical height for torus instability, $h_{\mathrm{crit}}=h(n_{\mathrm{crit}}=1.5)$, is the most robust measure to discriminate between ARs that host confined and eruptive flares. AR's exhibiting values of $h_{\mathrm{crit}}\lesssim40$~Mm above the flare-relevant PIL are likely to produce an eruptive flare. It segregates confined from eruptive flares better than alternative measures for the vertical decay of the magnetic field (the flux ratio), most likely due to its higher sensitivity to the nonlinear decay of the confining magnetic field above ARs.

3. The orientation of the magnetic field is organized distinctly different in the corona above ARs that host confined or eruptive flares. In ARs that host confined flares, the orientation of the flare-relevant PIL, $\Delta\varphi$, quickly adjusts to the underlying photospheric AR dipole field  with height. The adjustment ($\Delta\varphi\gtrsim40^\circ$) is completed already at the apex of the confining dipole field, i.e., at a height $h\simeq d_{\mathrm{PC}}/2$. That is different for ARs that host eruptive flares, where the flare-relevant PIL might be still considerably inclined near the apex of the underlying magnetic dipole field.

4. Though there exists a moderate degree of correlation between $\Delta\varphi$ and $h_{\mathrm{crit}}$, it appears that the critical height for torus instability is a stronger measure for the type of the upcoming flaring. This implies that the vertical decay of the confining field plays a stronger role than the particular orientation of the field at different heights. 

5.
The parameters deduced for ARs that host confined and eruptive events ($h_{\mathrm{crit}}$, flux ratio, $\Delta\varphi$) cannot be modeled to originate from a single normally distributed population of events, i.e., can be considered as to be distinctly related to the particular flare type.

\acknowledgments
\small
The authors acknowledge support from the Austrian Science Fund (FWF): P27292-N20. We are thankful for helpful discussions with F.\ P.\ Zuccarello. 
{ We acknowledge constructive comments from an anonymous referee and helpful suggestions from a statistics review.} 
\sdo\ data are courtesy of the NASA/\sdo\ AIA and HMI science teams. \sdo\ is a mission for NASA’s Living With a Star (LWS) Program. \goes\ is a joint effort of NASA and the National Oceanic and Atmospheric Administration (NOAA).

\normalsize
\bibliography{bibliography}  
\end{document}